\documentclass[fleqn,usenatbib]{mnras}

\usepackage{newtxtext,newtxmath}

\usepackage[T1]{fontenc}

\DeclareRobustCommand{\VAN}[3]{#2}
\let\VANthebibliography\thebibliography
\def\thebibliography{\DeclareRobustCommand{\VAN}[3]{##3}\VANthebibliography}

\usepackage{graphicx}	
\usepackage{amsmath}	

\title[cold streams in halos]{Evolution of cold streams in hot gaseous halos}

\author[W.S. Hong et al.]{
Wen-Sheng Hong,$^{1}$
Weishan Zhu,$^{2}$\thanks{E-mail: zhuwshan5@mail.sysu.edu.cn (WSZ)}
Tian-Rui Wang,$^{2}$
Xiaohu Yang,$^{1,3}$
Long-Long Feng $^{2}$
\\
$^{1}$Department of Astronomy, School of Physics and Astronomy, and Shanghai Key Laboratory for Particle Physics and Cosmology, \\~~~~Shanghai Jiao Tong University, Shanghai 200240, People's Republic of China\\
$^{2}$Department of Astronomy, School of Physics and Astronomy, Sun Yat-Sen University, No. 2 Daxue Road, Xiangzhou District, Zhuhai, 519082, China\\
$^{3}$Tsung-Dao Lee Institute, and Key Laboratory for Particle Physics, Astrophysics and Cosmology, Ministry of Education,\\ ~~~~Shanghai Jiao Tong University, Shanghai 200240, People's Republic of China
}

\date{Accepted XXX. Received YYY; in original form ZZZ}

\pubyear{2015}

\begin{document}
\label{firstpage}
\pagerange{\pageref{firstpage}--\pageref{lastpage}}
\maketitle

\begin{abstract}
In the prevailing model of galaxy formation and evolution, the process of gas accretion onto central galaxies undergoes a transition from cold-dominated to hot-dominated modes. This shift occurs when the mass of the parent dark matter halos exceeds a critical threshold known as $M_{shock}$. Moreover, cold gas usually flows onto central galaxies through filamentary structures, currently referred to as cold streams. However, the evolution of cold streams in halos with masses around $M_{shock}$, particularly how they are disrupted, remains unclear. To address this issue, we conduct a set of idealised hydrodynamic simulations. Our simulations show that (1) for a gas metallicity $Z=0.001-0.1Z_{\odot}$, cold stream with an inflow rate $\sim 3\, \rm{M_{\odot}}/yr$ per each can persist and effectively transport cold and cool gas to the central region ($< 0.2$ virial radius) in halos with mass $10^{12}\, \rm{M_{\odot}}$, but is disrupted at a radius around $0.2$ virial radius due to compression heating for halos with mass $3 \times 10^{12}\, \rm{M_{\odot}}$. (2) At $z\sim 2$, the maximum halo mass that capable of hosting and sustaining cold streams $M_{stream}$ is between $1\times 10^{12} \rm{M_{\odot}}$ and $1.5\times 10^{12}\rm{M_{\odot}}$ for gas metallicity $Z=0.001Z_{\odot}$, while for a higher gas metallicity $Z=0.1Z_{\odot}$, this value increases to $\sim 1.5\times 10^{12}\rm{M_{\odot}}$. (3) The evolution and ultimate fate of cold streams are determined primarily by the rivalry between radiative cooling and compression. Stronger heating due to compression in halos more massive than $M_{stream}$ can surpass cooling and heat the gas in cold streams to the hot ($\geq 10^6\,$ K) phase. 

\end{abstract}



\section{Introduction}
The acquisition of gas is a pivotal facet of galaxy formation and evolution, a process whose understanding has improved significantly over the past five decades. This knowledge has unfolded in tandem with the advancements in $\Lambda$CDM cosmology. Once dark matter halos are formed, they accrete gas from their cosmic surroundings. Following the epoch of reionization, completed at redshift $z\sim 5.5$, the gas that was initially acquired typically has a temperature of approximately $10^4$ K, the typical temperature of the intergalactic medium heated by the cosmic ultraviolet background (UVB). In the conventional scenario, the gas accreting onto halos will be heated to the virial temperature by shocks produced by the gravitational collapsing, then subsequently loses pressure support because of radiative cooling, and eventually settles into a rotation-supported disc, where stars were born (e.g., \citealt{1977MNRAS.179..541R, 1977ApJ...211..638S, 1978MNRAS.183..341W, 1980MNRAS.193..189F, 1985ApJS...58...39B, 1991ApJ...379...52W}). This picture was partly supported by hydrodynamical simulations conducted in the early 2000s, which had demonstrated that a substantial fraction of cold gas was indeed shock heated following its accretion (e.g., \citealt{2001MNRAS.327.1041B, 2003MNRAS.338..913H}). 

However, in a theoretical study on the collapse of protogalactic cloud, \cite{1977ApJ...215..483B} concluded that specific conditions related to the infalling velocity and the surface density of the descending gas are required to maintain adiabatic collapse shocks, which make the post-shock gas achieve virial temperature. Since the early 1990s, numerous simulations of galaxy formation with gas cooling have suggested that a substantial portion of the gas in galaxies has never been heated above $10^5$ K (e.g. \citealt{1991ApJ...377..365K, 2000MNRAS.316..374K, 2001ApJ...562..605F}). Later, \cite{2003MNRAS.345..349B} used an analytical model and one-dimensional simulations, revealing that virial shocks fail to develop in halos with masses below a critical threshold, approximately $10^{12}\, \rm{M_{\odot}}$ at a redshift of $z=2$ and are usually called as $\rm{M_{shock}}$, due to radiative cooling. Consequently, gas entering these less massive halos remains cold ($10^4-10^5$ K) along their journeys to the central region, usually called the cold accretion mode, and can drive a high star formation rate at high redshifts. Meanwhile, gas falling into halos more massive than the critical mass will be shock-heated to near the virial temperature. 

This scenario has been confirmed and further improved by several theoretical and simulation works in the past two decades. The transitions from cold-mode dominated accretion in low-mass halos and galaxies to hot-mode dominated accretion in massive systems have been explored in great detail (e.g. \citealt{2005MNRAS.363....2K, 2006MNRAS.368....2D, 2008MNRAS.390.1326O, 2009Natur.457..451D, 2009MNRAS.395..160K, 2011MNRAS.414.2458V, 2013MNRAS.429.3353N, 2021MNRAS.504.5702W}). The critical transition halo mass, $\rm{M_{shock}}$, exhibits modest variation in different studies, spanning a range from $10^{11.4}$ to $10^{12.0}\, M_{\odot}$ at a redshift of $z \sim 2$. Remarkably, simulations have illustrated that cold accretion predominantly occurs along filamentary structures and are often referred to as `cold streams'  (e.g., \citealt{2005MNRAS.363....2K, 2006MNRAS.368....2D, 2009Natur.457..451D, 2013MNRAS.429.3353N}). Furthermore, these cold streams can penetrate through shocks within halos with masses above $\rm{M_{shock}}$ at $z \gtrsim 1-2$, and the maximum halo mass that can host cold streams, denoted as $\rm{M_{stream}}$, increases with redshift (\citealt{2006MNRAS.368....2D}). 

Since the late 2000s, great efforts have been made to detect cold streams. Theoretical studies based on simulations suggest that identifying observationally cold streams through metal absorption lines is relatively difficult while the approach of Lyman-alpha absorption and emission is more promising (e.g. \citealt{2009MNRAS.400.1109D, 2010MNRAS.407..613G, 2011MNRAS.412L.118F,2011MNRAS.418.1796F, 2011MNRAS.413L..51K, 2012MNRAS.424.2292G}). However, despite several observed features that could potentially be attributed to cold accretion flows, there is still a lack of conclusive observational evidence for the existence of cold streams (e.g., \citealt{2013ApJ...776L..18C, 2014ApJ...786..106M, 2021A&A...649A..78D, 2021ApJ...908..188F}). More recently, \cite{2022ApJ...926L..21D} reports the finding of observational evidence for the transition from cold to hot accretion modes by measuring the ratio of Lyman-alpha luminosity from groups and clusters and the expected baryonic accretion rate at $z\sim 2$. Nevertheless, interpreting the observed features faces notable challenges and uncertainties. For instance, the Lyman-alpha emission may be partly powered by outflows and photoionization stemming from star formation and active galactic nuclei. On the other hand, the kinetics, clumpiness, and disrupted radius of cold streams remain somewhat unclear within the theoretical framework.

To consolidate the scenario of cold accretion through streams and to have a more robust interpretation of current and upcoming observations, it is imperative to dive into a deeper understanding of the evolution and fate of cold streams in massive halos. More specifically, several important open questions need to be addressed. How these cold streams are disrupted and lead to the closure of the cold accretion channel when the halo mass becomes comparable to $\rm{M_{stream}}$? How long can these cold streams survive against various perturbations and instabilities, including tidal forces due to interaction, Kelvin–Helmholtz, Rayleigh–Taylor and thermal instabilities (e.g. \citealt{2014MNRAS.439L..85W, 2016MNRAS.463.3921M, 2019MNRAS.489.3368B,2019MNRAS.484.1100M,2020MNRAS.494.2641M,ledos2023stability})? What impact does feedback from central galaxies and outflows have on the cold streams?

In this work, we utilize idealized three-dimensional hydrodynamic simulations to investigate the evolution of cold streams in isolated intermediate mass halos, aiming to gain insight into the mechanisms and processes of cold stream disruption in these halos. The layout of this paper is as follows: Section 2 describes the setup of our models for cold streams, hot gas halos, the simulations, and the typical timescales of relevant physical mechanisms. Detailed results from the simulations are presented in Section 3. In Section 4, we discuss the impact of the conduction process and the caveats of our study, as well as some factors that can be improved in future work. Finally, we summarise our findings in Section 5. 

\section{Methodology}
\label{sec:method}

\subsection{Models of hot gaseous halos and cold streams}
\label{sec:model}
\begin{figure}
    \centering
    \includegraphics[trim ={0.5cm 2cm 1cm 0cm},clip,width=1.0\columnwidth]{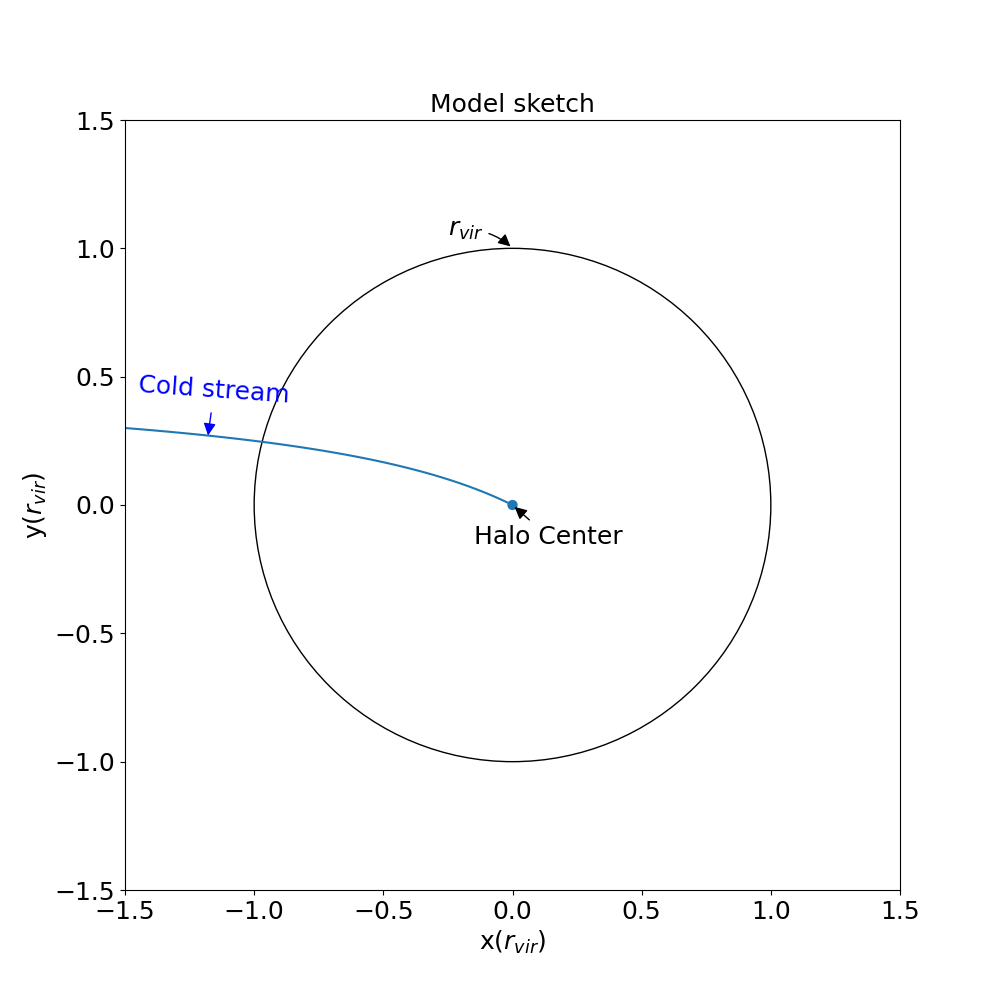}
    \caption{Schematic diagram of the model for cold stream and hot gaseous halo, showing a slice view of the central xy plane along the z direction. The simulation box has a cubic size of $(3r_{vir})^3$. Cold gas is injected into the box from the left boundary with an initial velocity along the x-axis, then will flow toward the halo center under the influence of static gravitational potential.}
    \label{fig: Schematic diagram}
\end{figure}

We adopt a simplified model of cold streams in the hot gaseous halo, which allows us to focus on the evolution of cold streams from the outside of the halo to the boundary of the central galaxy. We firstly set up a hot gaseous halo that is assumed to be in hydrostatic equilibrium under the gravitational potential of a dark matter halo with a given mass around the critical mass $\rm{M_{shock}}$, which indicates the gas accretion onto central galaxies transits from cold-mode dominated to hot-mode dominated at redshift $z=2$. The properties of the dark matter halo are assumed to be non-evolving during the simulation. Subsequently, we introduce a continuous injection of cold gas into the hot gaseous halo from a location beyond the virial radius. In the gravitational well, the cold gas flows towards the central region of the gaseous halo and experiences radiative cooling and compression along its streamline. Figure \ref{fig: Schematic diagram} provides a schematic diagram of our model. We track the evolution of cold streams in halos of different masses. The specifics of our models are outlined in the following subsections.

Before delving into the physics of our numerical models, it is essential to emphasise that our simplified model intentionally excludes the central galaxy and the associated physics of its interstellar medium (ISM), stellar component, and central supermassive black hole. Our primary objective is to concentrate on the evolutionary dynamics of the cold stream originating from external regions beyond the halo and extending downward to the boundary of the central galaxy. The latter has a radius around 0.10-0.15 times the virial radius of the halo. Consequently, we acknowledge that the results of our study possess limited predictive capacity in the inner region, given the absence of a detailed and realistic model in that particular domain.

\begin{table*}
    \centering
    \caption{Parameters for the hot gaseous halos and associated dark matter halos with different masses. The gaseous halos are assumed to be isothermal with a temperature $T_{h}$. The scale radius $R_{scale}$, concentration c, and scale density $\rho_{scale}$ are derived using the python package COLOSSUS with Planck18 cosmological parameters and ishiyama20 model at z=2. The gas density at the halo center $\rho_c$ is derived according to the cosmic baryon fraction (see Section 2.1 for more details).}
    \label{tab:Halo parameters}
    \begin{tabular}{|c|c|c|c|c|c|}
        \hline
         $M_{h}(\rm{M_{\odot}})$ & $T_{h}(\rm{K})$ & $R_{scale}(\rm{kpc})$ & c & $\rho_{scale}(\rm{g \times cm^{-3})}$ & $\rho_c(\rm{g \times cm^{-3}})$ \\
         \hline
         $1\times10^{12}$    & $3.30\times10^{6}$ & 31.12 & 3.90 & $3.35\times10^{-25}$ & $3.00\times10^{-24}$ \\
         $1.5\times10^{12}$  & $4.04\times10^{6}$ & 36.40 & 3.81 & $3.20\times10^{-25}$ & $3.995\times10^{-24}$ \\
         $3\times10^{12}$    & $5.71\times10^{6}$ & 47.27 & 3.70 & $2.99\times10^{-25}$ & $6.33\times10^{-24}$ \\
         \hline
    \end{tabular}
\end{table*}

\subsubsection{Hot Gaseous Halos}
\label{sec:hot-halo}
In our study, the background gravitational potential is determined by a dark matter halo that follows the Navarro-Frenk-White (NFW) profile (\citealt{1996ApJ...462..563N}),
\begin{equation}
\centering
    \begin{aligned}
    \rho_{dm}(r)&=\rho_{crit}\frac{\delta_{char}}{(r/R_{scale})(1+r/R_{scale})^2}\\
    &=\frac{\rho_{scale}}{(r/R_{scale})(1+r/R_{scale})^2},
    \label{eqn:dm_profile}
    \end{aligned}
\end{equation}
where $\delta_{char}$ and $\rho_{crit}$ are the characteristic overdensity parameter and the critical density of the universe, respectively; $\rho_{scale}$ denotes the scale density, and $R_{scale}$ is the scale radius. Consequently, the background gravitational potential can be expressed as,
\begin{center}
\begin{equation}
    \Phi(r)=-\frac{4\pi G\rho_{scale} R_{scale}^3}{r}ln(1+\frac{r}{R_{scale}})
\end{equation}
\end{center}

In this work, we consider dark matter halos with masses of $1.0, 1.5, $ and $ 3.0 \times 10^{12}\, \rm{M_{\odot}}$ at redshift $z=2$. The values of $R_{scale}$, $\rho_{scale}$ and c are calculated using the Python package COLOSSUS (\citealt{Diemer_2018}) with the Planck18 cosmological parameters (\citealt{2020A&A...641A...6P}), and the Ishiyama20 model (\citealt{Ishiyama_2021}). Under the static potential of the dark matter halos, the density profile of the associated gaseous halo under the hydrostatic equilibrium condition would be (\citealt{2010gfe..book.....M})
\begin{center}
\begin{equation}
    \rho_{gas}(r)=\rho_ce^{-b}(1+\frac{r}{R_{scale}})^{bR_{scale}/r},
\label{eqn:gas_profile}
\end{equation}
\end{center}
where $\rho_c=\rho(0)$ is the gas density at the halo center, and b is given by 
\begin{center}
\begin{equation}
    b=4\pi G\rho_{crit}\delta_{char}R_{scale}^2\frac{\mu m_p}{k_BT_h},
\end{equation}
\end{center}
where $m_p$ is the mass of a proton, $k_b$ is the Boltzmann constant, and $T_h$ is the temperature of halo gas. We determine the value of $\rho_c$ by assuming that the ratio of the gas mass to the dark matter mass within the virial radius is equal to the cosmological mean $\Omega_b/(\Omega_m-\Omega_b)$. The temperature of hot halo gas $T_{h}$ is set to around the virial temperature. The parameters relevant to the gaseous halo and their specific values are listed in Table.\ref{tab:Halo parameters}.  To reduce the possible boundary effect in our simulations, we will extend the gaseous halos to 1.5 times the virial radius $R_{vir}$, and the side length of the simulation box will be set to twice this enlarged scale. 

\subsubsection{Cold streams}
\label{sec:cold streams}

To mimic the cold streams, cold gas is injected continuously into the hot gaseous halo through a circular cross-section at the boundary region of the simulation box. Following the consideration in \cite{2014MNRAS.439L..85W}, we set the radius of the cross-section to be 15 kpc, which will have a covering factor close to previous studies (e.g., \citealt{2012MNRAS.424.2292G,2013ApJ...775...78F}). Under the influence of the gravitational field, the cold gas develops a conical stream entering the halo's central region. The injected cold gas is endowed with an initial velocity $v_s$ of 100 km/s along the x-direction at the boundary of the simulation box, which is close to the typical inflow velocity of cold flow at $r=1.5\,r_{vir}$ for halos with $M_h \sim 10^{12}\, \rm{M_{\odot}}$, that is, $\sim 0.7$ times of $v_{vir}=\sqrt{G M_h/r_{vir}}$, at redshift $z=2$ in cosmological simulation (\citealt{2015MNRAS.450.3359G}). The temperature of the cold streams is $3.3\times10^{4}$ K, which is moderately higher than the mean temperature of the IGM at $z=2$ that is heated by the cosmic UV background (see, e.g. \citealt{2021MNRAS.506.4389G}). However, it is comparable to the typical temperature of cold streams in cosmological hydrodynamic simulations (e.g., \citealt{2013MNRAS.429.3353N}). The mean density of cold gas is set to $3.0\times10^{-27} \rm{g\cdot cm^{-3}}$, which is an order of magnitude lower than that in \cite{2014MNRAS.439L..85W}. The inflow rate of a cold stream is 30 $\rm{M_{\odot}}$/yr in \cite{2014MNRAS.439L..85W}. An injection rate of 3-5 $\rm{M_{\odot}}$/yr should be more reasonable, as the cold accretion rate to halo is about 10-20 $\rm{M_{\odot}}$/yr at $z\sim 2$ for a dark matter halo with mass $ M_h \sim 10^{12}\, \rm{M_{\odot}}$ (e.g. \citealt{2011MNRAS.414.2458V, 2022ApJ...924..132Z}), and usually there are 3-4 streams in a single halo. For simplicity, the injection rate of each cold stream is fixed and does not depend on the halo mass. We will discuss the potential limitation of this assumption in Section 4. 

Note that previous simulations show that the gas in the cold streams is inhomogenous and partly clumpy (e.g., \citealt{2006MNRAS.368....2D,2008MNRAS.390.1326O,2013MNRAS.429.3353N}). Moreover, this inhomogeneity rises outside of the halos. Actually, the density of the intergalactic medium (IGM) indeed fluctuates to a certain extent. In the mildly non-linear regime, the probability distribution function (PDF) of the density of the IGM can be modelled by a log-normal random field (\citealt{1997ApJ...479..523B}). In the highly non-linear regime, the IGM exhibits turbulent features (\citealt{2013ApJ...777...48Z,2017ApJ...838...21Z}), and shows an extended density distribution. Taking into account these factors, we further assume that the gas density of the injected cold gas follows a log-normal distribution of supersonic turbulence (e.g., \cite{Konstandin_2016}). Specifically, the mean of the logarithmic density is denoted as s = $ln(\rho)$ and is given by :
\begin{equation}
    \mu_{s} = \ln(\mu_{\rho}) - \sigma^2_s/2
\end{equation}
in which $\mu_s, \mu_{\rho}$ are the mean values of the natural logarithm of the mass density and the mean of the original mass density, respectively. The standard deviation of $\sigma_s$ is related to the Mach number as follows,
\begin{equation}
    \sigma_s^2 = \ln(1+b_M^2 \mathcal{M}^2),
\end{equation}
where $b_M$ can be interpreted as the ratio of the compressive Mach number to the total root-mean-square Mach number of the flow, i.e.
\begin{equation}
    b_M = \frac{\mathcal{M}_c}{\mathcal{M}}
\end{equation}
The value of $b_M$ depends on the driving force of turbulence. For purely solenoidal forcing, $b_M$ = 1/3, and for purely compressive forcing, $b_M$ = 1. In this work, we adopt $b_M = 0.5$, since both solenoidal and compressive modes are presented in the turbulent IGM (\citealt{2013ApJ...777...48Z}).

\begin{table}
    \centering
    \caption{Parameters for the cold stream: It has a log-normal density distribution with mean value $\rho_s$, temperature $T_s$, and an initial velocity of $v_s = $100 km/s. The radius of the cross-section at the initial position is 15 kpc, }
    \label{tab:Cold stream parameters}
    \begin{tabular}{|c|c|c|c|}
        \hline
         $T_{s}(\rm{K})$ &  $\rho_{s}(\rm{g \times cm^{-3})}$ & $R_{s}(\rm{kpc})$ & $v_s$(km/s) \\
         \hline
         $3.3\times10^{4}$  & $3.0\times10^{-27}$ & 15 &  100km \\
         \hline
    \end{tabular}
\end{table}

\subsection{Simulations}
\label{sec:simulations}

To investigate the evolution of cold streams in halos, we run hydrodynamic simulations using the ATHENA++ code (\citealt{Stone2020}). The simulations are performed in a 3D Cartesian geometry with a uniform cubic grid. The simulation box has a side length of 3$r_{vir}$, and the halo center is located at the origin (0,0,0). The cold stream is injected at a specific position (-1.5$r_{vir}$, 0.3$r_{vir}$, 0), taking into account that cold streams could have a non-negligible angular moment.

Several relevant physical processes may affect the evolution of cold streams, including cooling, conduction, and the strength of the gravitational potential. These factors were considered in our simulations to identify the mechanisms that lead to the disruption of cold streams. In addition, the impacts of the cold streams' density perturbation and the resolution of the simulation were also explored. We have run a set of simulations to investigate the roles of each factor by turning on/off different processes or modifying relevant parameters in different simulations. Hereafter, the simulations are named `X(X.Y)e12-IJK-abc(-HZ)', where `X(X.Y)e12' represents the mass of host dark matter halo, `IJK' is the number of grids along each box side, the first character in `abc' indicates whether there is a perturbation in the density of cold streams or not, denoted as `p' or `n' respectively. The second (third) character in `abc' stands for whether cooling (conduction) is turned on or off, denoted by `c', or `n' correspondingly. `HZ' indicate a gas metallicity of $Z = 0.1Z_{\odot}$, while other simulations adopt a metallicity of $Z = 0.001Z_{\odot}$. The full list of our simulations can be found in Table \ref{tab: Simulations Groups}. More details about the simulations are described below. 

\begin{table}
    \centering
    \caption{List of all the simulations performed, the first and second columns are the serial number and gas metallicity, respectively. The labels in the third column indicate the dark halo mass, resolution, and other settings (see section \ref{sec:simulations} for more details).}
    \label{tab: Simulations Groups}
    \begin{tabular}{|c|c|c|}
         \hline 
         No. & Metallicity($Z_{\odot}$) & Mass($M_{\odot}$)-Resolution-Condition(-other condition)  \\
         \hline 
         1 & 0.1 & 1e12-256-pcn-HZ \\
         2 & 0.1 & 1.5e12-256-pcn-HZ \\
         3 & 0.1 & 3e12-256-pcn-HZ \\
         4 & 0.001 & 1e12-256-ncn \\
         5 & 0.001 & 1e12-256-pcn \\
         6 & 0.001 & 1e12-256-pnc \\
         7 & 0.001 & 1e12-256-pcc \\
         8 & 0.001 & 1e12-512-pcn \\
         9 & 0.001 & 1.5e12-256-pcn \\
         10 & 0.001 & 3e12-256-pcn \\
         11 & 0.001 & 1e12-256-xcn-noinflow \\
         12 & 0.001 & 3e12-256-xcn-noinflow \\
         \hline \\
    \end{tabular}
\end{table}

\subsubsection{Radiative Cooling and thermal conduction:}
\label{sec::cooling and conduction}
In the simulations, the effects of radiative cooling and thermal conduction have been considered, which allows us to analyse how they may affect the fate of cold streams. We have implemented these two effects by solving the following hydrodynamic equations with the Athena++ code (\citealt{Stone2020}):

\begin{equation}\label{eqn:energy equation}
	\left\{
	\begin{aligned}
		&\frac{\partial \rho}{\partial t} + \triangledown \cdot(\rho \textbf{v}) = 0, \\
            &\frac{\partial (\rho \textbf{v})}{\partial t} + \triangledown\cdot[\rho(\textbf{v}\otimes \textbf{v}) + P] = 0 \\
		&\frac{\partial e}{\partial t} + \triangledown\cdot [(e+P)\textbf{v}] = -\Lambda_{net} - \triangledown\cdot\textbf{q} 
	\end{aligned}
	\right.
\end{equation}

where $\rho$ and $\textbf{v}$ stand for the density and velocity respectively, $e$ represents the energy density, $P = (\gamma - 1)U$ denotes the pressure, $U$ signifies the internal energy density, and $\gamma = 5/3$; $\Lambda_{net}=n^2\Lambda_{norm}$ stands for the net cooling rate per unit volume, where $n$ is the number density and $\Lambda_{norm}$ is the normalised cooling function and is obtained from the cooling table given by \cite{1993ApJS...88..253S} in this work;$\mathbf{q}$ is the conduction flux. Cooling and thermal conduction are handled by updating the energy density term $e$ after solving hydrodynamic equations without source terms at each time step. That is, $\Lambda_{net}$ and $\mathbf{\nabla}\cdot\mathbf{q}$ are treated as source terms in the equation of energy density. 

The cooling rate and conduction flux in each cell at each time step are calculated using the methods described below.

\begin{figure}
    \centering
    \includegraphics[trim ={0.5cm 0.3cm 1cm 0cm},clip,width=\columnwidth]{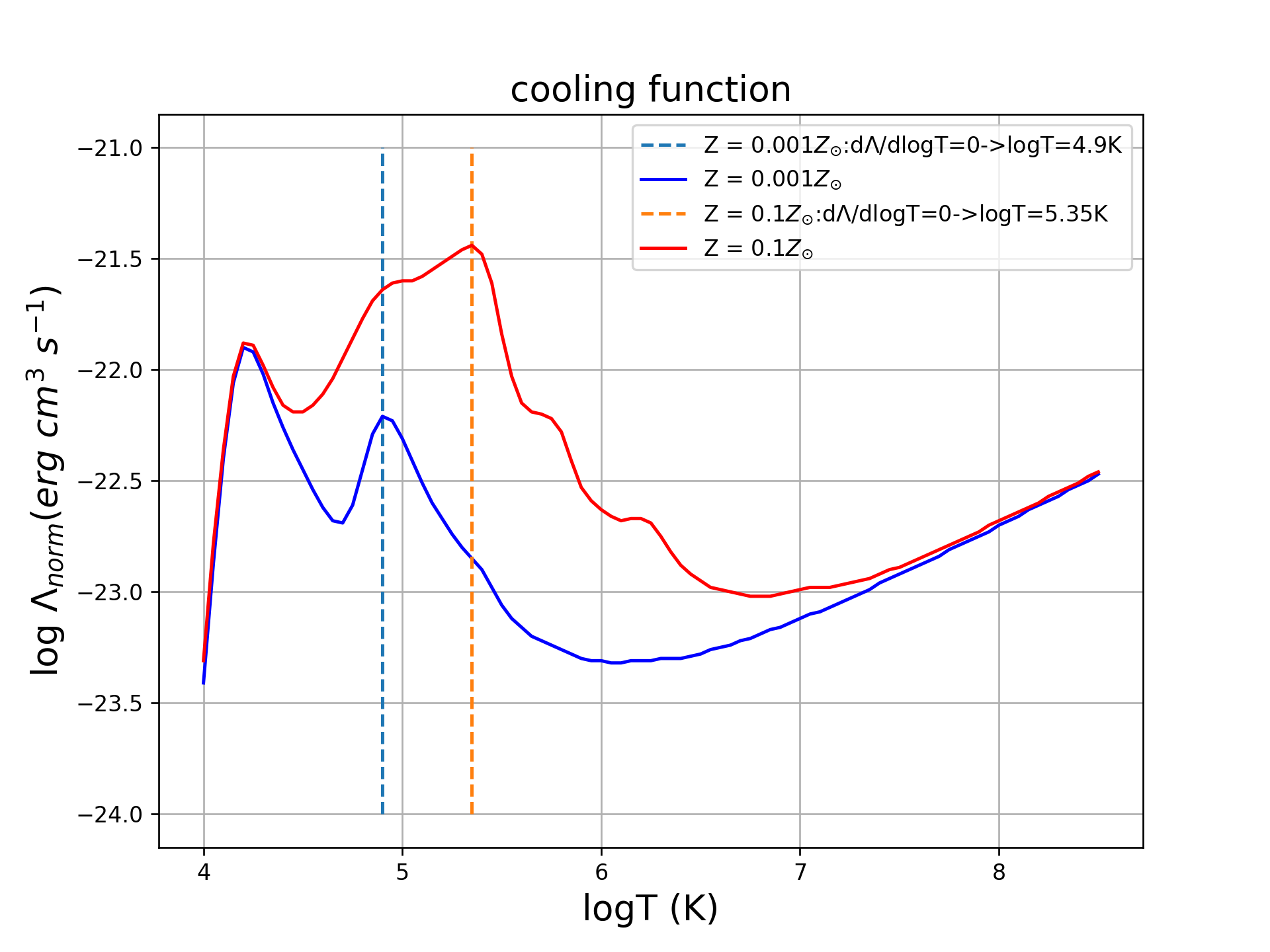}   
    \caption{Normalised cooling rate as a function of the gas temperature for metallicity Z=0.1$Z_{\odot}$(red line) and Z=0.001$Z_{\odot}$(blue line). The vertical dotted lines indicate the local maximum around $10^5$ K for both metallicity, i.e., $T = 10^{4.9}\,\rm{K}$ for $Z = 0.001Z_{\odot} $ and $T = 10^{5.35}\,\rm{K}$ for $Z = 0.1Z_{\odot} $ respectively.}
    \label{fig:coolingfunction}
\end{figure}

The cooling rate of the gas in a grid cell depends on its temperature, hydrogen number density, and metallicity. In this work, we assume that the cold streams and the gaseous halo have the same metallicity for simplicity. Yet, to explore the impact of metallicity, we adopt two different metallicities within our simulations, i.e., $Z = 0.001Z_{\odot}$, and $Z = 0.1Z_{\odot}$, respectively. The former value is adopted because the metallicity of the diffuse IGM at $z \sim 2$ is estimated to be $10^{-2}-10^{-3} Z_{\odot} $ (e.g., \citealt{2008ApJ...689..851A, 2009MNRAS.399..574W,2016MNRAS.463.2690D}). The latter value is somewhat lower than the metallicity of the ISM at $\sim 2$, i.e., $0.16-1.0 Z_{\odot}$, (e.g., \citealt{2010ApJ...719.1168E, 2014ApJ...795..165S,2019MNRAS.484.5587T,2023ApJ...942...24S}). Thus, the metallicity of gas in the cold stream at $z \sim 2$ is very likely between $Z = 0.001Z_{\odot}$ and $Z = 0.1Z_{\odot} $. For each simulation with a given metallicity, we use the cooling table of the corresponding metallicity provided by \cite{1993ApJS...88..253S} to find the normalised cooling rate, $\Lambda_{norm}$, of gas in different cells by interpolation, where the temperature of the gas is inferred from the internal energy and density.
The normalised cooling rate, $\Lambda_{norm}$, as a function of gas temperature for these two metallicities, is shown in Fig. \ref{fig:coolingfunction}. The vertical dotted lines mark the peak of the cooling rates at around $10^5 $ K. If the gas temperature approaches the corresponding value for a particular metallicity, the cooling rate is boosted, which could cause the gas to cool down rapidly if the increment of pressure due to compressional heating is insufficient. 

The cooling function employed in our study takes into account the cooling of gas with a temperature higher than $10^4$ K. That is, cooling below temperature $10^4$ K is totally neglected. We expect that such an implementation has a minor effect on our main results about the evolution of the cold stream from somewhere outside of the halo inward to the boundary of the central galaxy because of the following considerations. First, the normalised cooling rate reduces significantly from around $10^{-22.5}$ to $10^{-24.5}\, \rm{erg\,s^{-1}\,cm^3}$ for a gas number density of $1/\rm{cm^3}$, when the temperature drops from slightly above $10^4$ K to several $10^3$ K (e.g. \citealt{2007MNRAS.379..963M,2017MNRAS.466.2217S}). Second, molecular cooling is the primary cooling mechanism of gas below $10^4$ K. It usually requires a total hydrogen number density of $n_H>10/\rm{cm^3}$ to form molecules effectively (e.g.,\citealt{2010MNRAS.404....2G,2017MolAs...9....1W}), which is much higher than the typical number density of gas in the cold stream, i.e., $\sim 10^{-3}-10^{-1}/\rm{cm^3}$.

On the other hand, radiative heating from the cosmic UVB and central galaxy is not included in our simulation. The initial temperature adopted for gas in the cold stream has accounted for the heating of IGM by the UVB. Moreover, self-shielding would partially attenuate the heating of UVB. In addition, neglecting the cooling below $10^4$ K can partially compensate for the absence of UVB. Yet, the omission of radiative heating from the central galaxy, including UV and X-rays from accretion to the SMBH, would lead to over-cooling for the hot gas in the central region. \cite{2020MNRAS.494.2641M} shut off cooling for gas with T >$\sim\ 2\times 10^5$ K to prevent the hot gas from over-cooling after long periods. To compensate for the probable overcooling of the hot gas in the central region in our simulations, we manually reduce the cooling rate, $\Lambda_{\rm{net}}$, of the gas at each simulation step such that the cooling time is not shorter than 1Gyr for gas cells with a temperature of $\rm{T} > \ 5\times 10^5$ K and a number density of $n> 1\times 10^{-2} \rm{cm}^{-3}$. In practice, this remedy works only for gas within the inner 0.2 $r_{vir}$ of halos in our simulations.

In addition to radiative cooling, thermal conduction, i.e., the transfer of energy due to temperature gradients, has also been found to play an important role in the survival of cold clouds within ambient gas (e.g., \citealt{2016MNRAS.462.4157A}; \citealt{Armillotta_2017}). Thermal conduction may affect the cooling of hot gas and the diffusion of cold streams, especially in the diffuse and intermediate stream regimes (\citealt{ledos2023stability}). The second term on the RHS of equation \ref{eqn:energy equation} represents the thermal conduction process.
According to the Spitzer formula (\citealt{Spitzer1962}), the heat conduction flux is
\begin{equation}
    \centering
    q = - \kappa_{sp}\triangledown T , 
\end{equation}
where $\triangledown T$ is the temperature gradient, and the heat conduction coefficient is:
\begin{equation}
    \centering
    \kappa_{sp} = \frac{1.84\times 10^{-5}}{\ln\Psi}T^{5/2}  \, \rm{erg\ s^{-1} K^{-1} cm^{-1}} , 
\end{equation}
where $ln\Psi$ is the Coulomb logarithm:
\begin{equation}
    \centering
    ln\Phi =  29.7 + \ln[\frac{T/10^6\rm{K}}{\sqrt{n/\rm{cm}^{-3}}}]
\end{equation}
Following the consideration and treatment in \cite{Armillotta_2017}, the heat conduction flux can be modified to:
\begin{equation}
    \centering
    \label{equ:kappa}
    q = -0.1 \frac{\kappa_{sp}}{1+\sigma}\triangledown T
\end{equation}
where $\sigma$ indicates the absolute ratio of the classical heat flux to the saturated heat flux. In this work, we use $\sigma=49$, because we find that such a value can reproduce the results well in \cite{Armillotta_2017}.

Before moving on to more details of the simulations, we first estimate the relative strength between cooling and thermal conduction. In this regard, comparing the Field length (\citealt{1990ApJ...358..375B}; \citealt{2016MNRAS.462.4157A}), defined by
the maximum length scale on which heat energy transport is effective, with the stream radius, can serve as a valuable indicator. The Field length read as,

\begin{equation}
    L_{field} = (\frac{\kappa T_h}{n_s^2\Lambda(T_s)})^{1/2}
\end{equation}

where $\kappa=0.1\frac{\kappa_{sp}}{1+\sigma}$ represents the heat conduction coefficient introduced above, $n_s,T_s$ is the number density and temperature of the cold stream, respectively, and $T_h$ is the hot halo temperature. The function $\Lambda(T_s)$ represents the cooling rate with temperature $T_s$. In our models, the Field length is 10 (20) pc for halo mass $1.0 \times \, 10^{12} \rm{M_\odot}$ ($3.0 \times \, 10^{12} \rm{M_\odot}$) and metallicity Z=$0.001Z_{\odot}$, which is much smaller than the typical radius of a stream, i.e., 2-15 kpc. Even if the value of $\sigma$ is reduced to 2-5, the Field length is still far below the radius of the stream. Moreover, as we will see, the typical cooling timescale is much shorter than the thermal conduction timescale. On the basis of these results, we expect that thermal conduction is unlikely to affect our primary conclusions regarding the evolution of cold streams in the presence of radiative cooling.  

Nevertheless, to provide a comprehensive evaluation, we will inspect the effect of thermal conduction on the evolution of cold streams by turning the thermal conduction on and off in different simulations. 

\subsubsection{Kelvin Helmholtz Instability (KHI) and comparison of multi-time scales:}

Some previous studies suggest that Kelvin-Helmholtz instability(KHI) can play a significant role in the evolution of the stream (e.g., \citealt{2016MNRAS.463.3921M,2019MNRAS.484.1100M,2020MNRAS.494.2641M}). For a better understanding of our simulations, it is beneficial to evaluate the probable overall effects of KHI.  If radiative cooling is included, a cold stream with a radius greater than the critical radius $R_s$(\citealt{Aung_2019,2020MNRAS.494.2641M}) would grow in mass, instead of being disrupted by the KHI.
The critical stream radius $R_s$ is given by \cite{2020MNRAS.494.2641M}, 

\begin{equation}
    R_{s,crit} \simeq 0.3\, \rm{kpc}\ \alpha_{0.1} \delta_{100}^{3/2}\mathcal{M}_{h}\frac{T_{4}}{n_{s,0.01}\Lambda_{mix,-22.5}}
\end{equation}

where $\alpha_{0.1} = \alpha/0.1 = 0.21\times(0.8exp(-\mathcal{M}_{tot}^2)+0.2)/0.1$, $\mathcal{M}_{tot} = v_s/(c_s+c_{h})$ is the total Mach number, $c_s$ and $c_{h}$ denote the sound speed of the stream and the background halo gas, respectively. $\delta_{100}=\delta/100$, where $\delta=\rho_{s}/\rho_{h}$ represents the ratio of the density of gas in the stream to the density of the background halo gas. $\mathcal{M}_{h} = v_s/c_{h}$ represents the Mach number of background halo gas. Furthermore, $T_4=T_s/(10^4\,\rm{K})$ indicates the temperature of the stream in units of $10^4$ K. Additionally, $n_{s,0.01} = n_s/0.01\ \rm{cm^{-3}}$ and $\Lambda_{mix,-22.5} = \Lambda_{mix}/10^{-22.5}\ \rm{erg\ s^{-1}\ cm^{3}}$. The subscript `mix' indicates the turbulent mixing layer (\citealt{1990MNRAS.244P..26B}) formed by the turbulent mixing driven by KHI at the interface of two-phase gas: cold, dense gas, and hot, diffuse gas. Recent studies(\citealt{Gronke2018TheGA}; \citealt{2020MNRAS.494.2641M}; \citealt{Chen_2023}) have revealed that cooling in these layers can significantly impact the fate of the cold streams. In our model, the critical stream radius $R_{s,crit}$ is typically 1.54 (7.69 kpc) at r=1.0 $R_{vir}$ for a halo mass of $1.0 \times \, 10^{12}\, \rm{M_\odot}$ ($3.0 \times \, 10^{12}\, \rm{M_\odot}$) and a metallicity Z=$0.001Z_{\odot}$. These results are detailed in Table.\ref{tab: timescale} and are smaller than the original radius of the cold stream. As shown in \cite{2020MNRAS.494.2641M}, $R_{s,crit}$ would decrease with r in a trend similar to $R_{s}$. Therefore, the condition $R_{s,crit}<R_{s}$ would hold throughout the evolution of a cold stream. According to \cite{2020MNRAS.494.2641M}, it is reasonable to expect that the KHI would not dissolve the stream in our model.

Alternatively, the collective impact of various mechanisms influencing the evolution of a cold stream such as Kelvin-Helmholtz Instability (KHI), cooling, thermal conduction, and compressional heating can be estimated by considering their respective typical timescales. For a 3-dimensional cylindrical cold stream, \cite{2016MNRAS.463.3921M} and \cite{2019MNRAS.484.1100M} show that the characteristic time for the growth of perturbations with wavelengths $\lambda \sim 1-10\,R_s$ due to KHI in the linear regime, $t_{\rm{KH}}$, is related to the sound crossing timescale, $t_{sc}$, as:

\begin{equation}
    t_{\rm{KH}} \simeq \,(0.5 \sim 1.0)\  t_{sc} \simeq\  (0.5 \sim 1.0)\ \frac{2R_{s}}{c_s}.
\end{equation}

It suggests a $t_{\rm{KH}}$ of around 0.9-1.8 Gyr for the cold stream in our model, i.e., about 1/4 to 1/2 of the simulation time we set. 

The thermal conduction timescale for the stream can be derived from the analysis in \cite{ledos2023stability},

\begin{equation}
    \tau_{s} = \frac{n_sk_BR_{s}^2}{\kappa(\gamma-1)}.
\end{equation}


It gives a thermal conduction timescale of up to $10^4$ Gyr for our model, which is significantly larger than other timescales.

The cooling timescale reads as, 
\begin{equation}
       t_{cool} = \frac{k_BT}{(\gamma-1)n\Lambda_{net}}.\\ 
\end{equation}

We have evaluated both the cooling time of the gas in the hot halo and in the mixing layer at one and a half virial radius for both halos with mass $1.0 \times \, 10^{12} \rm{M_\odot}$ and $3.0 \times \, 10^{12} \rm{M_\odot}$, respectively. Both two gas metallicity Z=$0.001Z_{\odot}$ and Z=$0.1Z_{\odot}$ are considered. As shown in \citealt{2020MNRAS.494.2641M}, the mean number density
and temperature of the mixing layer to be $n_{mix} \simeq (n_s\times n_{h})^{1/2}$ and $T_{mix} \simeq (T_s\times T_{h})^{1/2}$ respectively. The values of estimated cooling timescales are listed in Table \ref{tab: timescale}. For the hot halo gas, the cooling time is much longer than the Hubble time at $r=r_{vir}$, and drops to several Gyr at $r=0.5\,r_{vir}$. In contrast, the cooling of gas in the mixing layer is much more effective. The corresponding cooling time scale is about 1-3 Gyr at $r=r_{vir}$ for Z=$0.001Z_{\odot}$, and decreases to less than 0.3 Gyr for Z=$0.1Z_{\odot}$. The cooling time of gas in the mixing layer decreases sharply as r decreases. 

\cite{2006MNRAS.368....2D} demonstrated that if the cooling rate exceeds the compression rate required to restore pressure in the post-shock gas, the post-shock gas will become unstable against radial gravitational contraction and cannot sustain the shock. Conversely, when the cooling rate is slower, the pressure acquired through compression can counterbalance the loss from radiative cooling, allowing the post-shock gas to remain stable against global gravitational collapse and thus support the shock. Specifically, the timescale of compressional heating is given by

\begin{equation}
    t_{comp} = \frac{\Gamma}{(-\triangledown\cdot \textbf{u})}
\end{equation}

where $\Gamma \equiv \frac{3\gamma+2}{\gamma(3\gamma-4)}$. The compressional heating timescales in all of our simulations are approximately 0.5 Gyr, which is shorter than the cooling timescale of hot halo gas in most regions. For the gas in the mixing layer, the cooling timescales are larger or comparable to the compression timescale for gas metallicity Z=$0.001Z_{\odot}$ at $r>r_{vir}$ but are shorter for Z=$0.1Z_{\odot}$. 

Given the timescales estimated above, compressional heating is expected to exert a more significant influence than Kelvin-Helmholtz instability (KHI) on the evolution of the cold stream in all the cases considered and dominate the evolution for the case Z=$0.001Z_{\odot}$ and $M_h=3.0 \times \, 10^{12} \rm{M_\odot}$. However, cooling is expected to become more important if the metallicity is increased or the halo is less massive and will be a dominant player in case Z =$0.1Z_{\odot}$ and $M_h=1.0 \times \, 10^{12} \rm{M_\odot}$. Thermal conduction would play a limited role in our models.

\begin{table*}
    \centering
    \caption{This table displays cooling timescales and typical critical radius of gas in the hot gas and in the mixing layer. The first and second columns represent the gas metallicity and the halo mass. The third indicates the distance from the center of the halo. The fourth, fifth, and sixth columns are the temperature, number density, and cooling timescales of hot halo gas, respectively. The seventh to ninth columns are the same as the fourth to sixth but for the gas in the mixing layer. The last column is the typical critical radius.}
    \label{tab: timescale}
    \begin{tabular}{|c|c|c|c|c|c|c|c|c|c|}
         \hline 
         Metallicity($Z_{\odot}$) & Mass($\rm{M_{\odot}}$) & r($r_{vir}$) &T(K) & n($\rm{cm^{-3}}$) &$t_{cool}$(Gyr) &$T_{mix}$(K) & $n_{mix}$(\rm{$cm^{-3}$}) & $t_{cool,mix}$(Gyr) &  $R_{s,crit}(\rm{kpc})$\\
         \hline 
         0.001 & $1\times 10^{12}$ & 1.0 & 3.3$\times 10^6$ & 6.1$\times 10^{-5}$ &67 & 3.3$\times 10^5$ & 2.5$\times 10^{-4}$ & 1.05 & 1.54\\
         0.001 & $3\times 10^{12}$ & 1.0 & 5.7$\times 10^6$ & 2.5$\times 10^{-5}$ &244 & 4.3$\times 10^5$ & 1.6$\times 10^{-4}$ & 2.76 & 7.69\\
         0.001 & $1\times 10^{12}$ & 0.5 & 3.3$\times 10^6$ & 6.8$\times 10^{-4}$ &6 & 3.3$\times 10^5$ & 8.4$\times 10^{-4}$ & 0.29 & 0.04\\
         0.001 & $3\times 10^{12}$ & 0.5 & 5.7$\times 10^6$ & 4.5$\times 10^{-4}$ &13 & 4.3$\times 10^5$ & 6.8$\times 10^{-4}$ & 0.65 & 0.09\\
         0.1 & $1\times 10^{12}$ & 1.0 & 3.3$\times 10^6$ & 6.1$\times 10^{-5}$ &32 & 3.3$\times 10^5$ & 2.5$\times 10^{-4}$ & 0.07 & 0.10\\
         0.1 & $3\times 10^{12}$ & 1.0 & 5.7$\times 10^6$ & 2.5$\times 10^{-5}$ &159 & 4.3$\times 10^5$ & 1.6$\times 10^{-4}$ & 0.27 & 0.67 \\
         0.1 & $1\times 10^{12}$ & 0.5 & 3.3$\times 10^6$ & 6.8$\times 10^{-4}$ &3 & 3.3$\times 10^5$ & 8.4$\times 10^{-4}$ & 0.02 &0.002\\
         0.1 & $3\times 10^{12}$ & 0.5 & 5.7$\times 10^6$ & 4.5$\times 10^{-4}$ &9 & 4.3$\times 10^5$ & 6.8$\times 10^{-4}$ & 0.06 & 0.009 \\
         \hline
    \end{tabular}
\end{table*}

\subsubsection{Tracer of cold streams}

To trace the cold gas injected from the boundary, we use the passive scalar module in Athena++. Based on the tracer field, we define a quantity called `$w_{trac}$', which is the ratio of the density of gas that was injected as a cold stream, denoted as $\rho_{stream}$, to the total gas density $\rho_{gas}$, in each grid cell. More specifically, $w_{trac}$ at a grid (i,j k) reads as,
\begin{equation}
    w_{trac}(i,j,k) = \frac{\rho_{stream}(i,j,k)}{\rho_{gas}(i,j,k)},
\end{equation}
where $i,j,k$ are the ranks of the grid along three dimensions. If $w_{trac}$ equals one at a particular grid, this grid cell is entirely occupied by gas injected as part of a cold stream at the boundary. When $w_{trac}$ equals 0,  the cold stream has not yet reached the corresponding grid cell.

\subsubsection{Resolution}
\label{sec:resolution}

To inspect the impact of resolution, we have run two simulations that track the evolution of a cold stream in a halo with mass $10^{12} \rm{M_\odot}$, using two different resolutions, $512^3$ and $256^3$ grids, respectively. Except for the resolution, all the other settings are the same in these two simulations, `1e12-256-pcn' and `1e12-512-pcn'. A gas metallicity of $Z=0.001Z_{\odot}$ is adopted. We find that the overall results are similar between these two resolutions. However, the accumulation of cold gas in the inner region ($r<0.2r_{vir}$) starts earlier in the higher resolution run (see Appendix \ref{sec:appd-resolution} for more details). However, the central galaxy is omitted from our current model. A realistic model of the central galaxy is required to obtain more reliable results on gas evolution in the central region, in addition to higher resolution.

As claimed in \cite{Nelson_2020}, a resolution finer than $\Delta x$ < 100 pc is necessary to fully resolve the small-scale structure of cold gas in the CGM. Consequently, a simulation with a box size of ~300 kpc requires nearly 3000 grids in each dimension for a uniform grid, which is extremely costly for a 3D hydrodynamic simulation. Adaptive mesh refinement would be helpful in resolving this issue. However, designing the refinement strategy for the evolution of the cold stream in the CGM is not straightforward. On the other hand, the timescales described in the previous section indicate that the dominant process in our models would be cooling and compressional heating. Consequently, to optimize computational resources, we have performed additional simulations utilizing a default configuration of $256^3$ fixed grids, i.e., reaching a resolution around $1.2$ kpc. The default resolution we adopted may have somewhat underestimated the impact of instabilities on scales of $\sim 100$ pc; however, it is expected to adequately capture the overall feature of stream evolution at radial distances $r>0.2r_{vir}$, concluded from a comparative analysis between the simulations `1e12-256-pcn' and `1e12-512-pcn'.

\subsubsection{Boundary condition}

In our model, the gravitational field is assumed to be static. We found that the outflow boundary conditions built into Athena++ are not suitable for our simulation, as they can cause instability in the ghost zone at the boundaries. To address this issue, we have implemented user-defined boundary conditions. Specifically, we design the boundary condition in those ghost zones adjacent to the stream to allow cold streams to continuously flow into the halo, while the properties of the gas in the other ghost zones are set to fix to the profile of the gaseous halo as described in the section \ref{sec:model}. Under this setting, the inflow and outflow at the boundary adjacent to hot halo gas is somewhat suppressed. Nevertheless, the hot halo can sustain a steady state throughout the simulation. 

\subsubsection{Simulations without inflow cold streams}

\begin{figure}
    \centering
    \includegraphics[trim ={0.5cm 1cm 1.5cm 1.0cm},clip,width=\columnwidth]{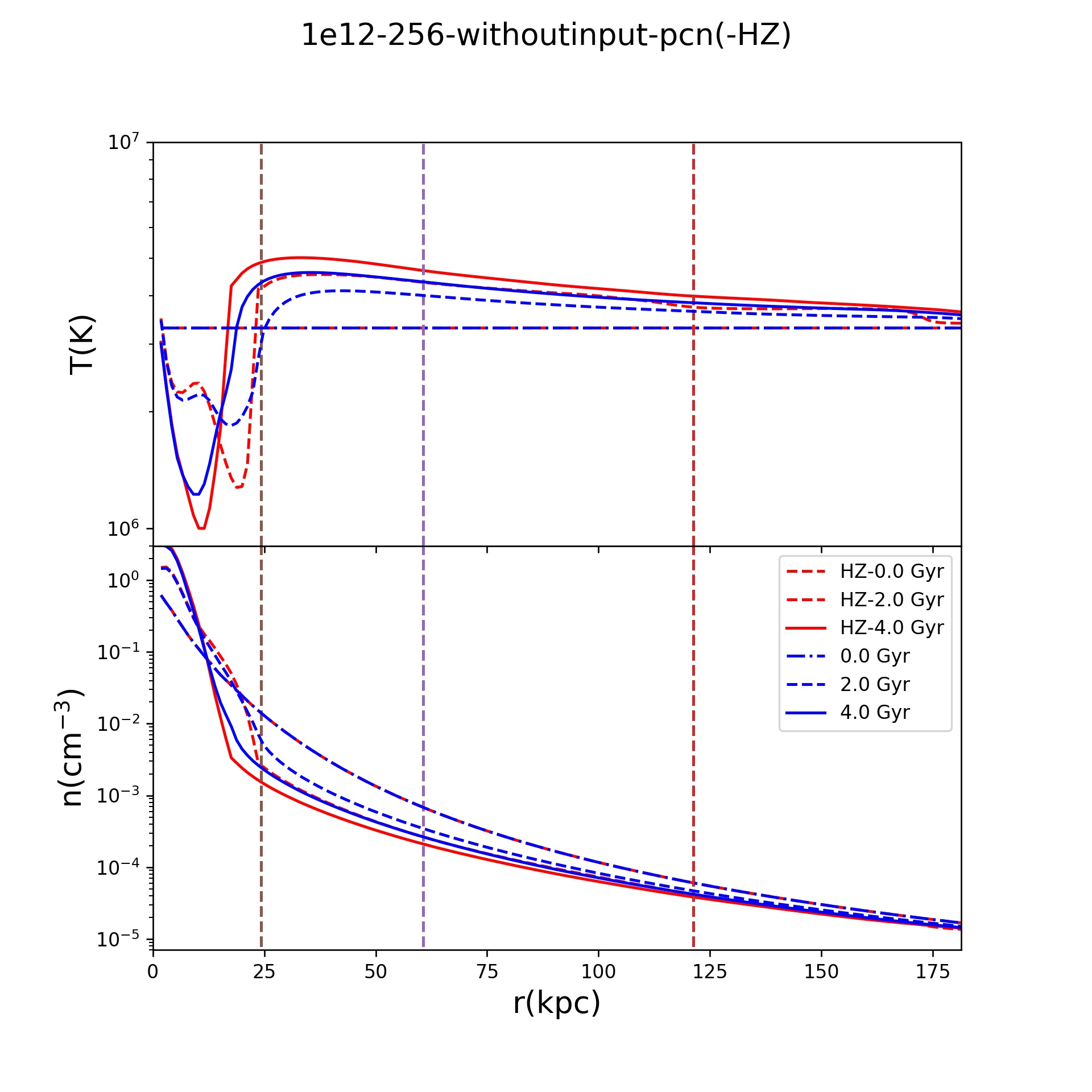} 
    \caption{The average temperature (top) and density (bottom) of the gas at the initial time t=0.0 Gyr (dash-dotted lines),2.0 Gyr(dashed line) the final moment 4Gyr (solid lines) for simulation 1e12-256-xcn-noinflow, i.e., without injected cold streams, with metallicities $Z=0.1Z_{\odot}$(red) and $0.001Z_{\odot}$(blue)}
    \label{fig:without-inflow}
\end{figure}

The density profile of the gaseous halo given by equation \ref{eqn:gas_profile} is derived without considering cooling. Before we explore the evolution of the cold stream in the halo, it is necessary to check the evolution of the hot gaseous halo under the influence of cooling and the gravitational potential. To this end, we have performed four test simulations without injected cold streams for halo mass $1.0$ and $3.0 \times \, 10^{12} \rm{M_\odot}$, with metallicity 0.001$Z_{\odot}$ and 0.1$Z_{\odot}$ respectively. The temperature and density profiles as a function of radius at time 0, 2.0, and 4.0 Gyr for the simulations with a halo mass $M_h = 1.0 \times \, 10^{12} \rm{M_\odot}$ are depicted in Figure \ref{fig:without-inflow} for both metallicities. After 4 Gyr, the temperature and number density of the halo have experienced notable changes. The change in the inner region is more violent, showing an evident temperature decrease and an increment of the density gradient. On the other hand, the gas temperature is moderately enhanced while the density decreases in the outer region. The results of the other two simulations with $M_h = 3.0 \times \, 10^{12} \rm{M_\odot}$ are similar (Figures are not shown, for the sake of conciseness).

These changes can be explained by the combined effect of cooling and compressional heating. The cooling time of the hot halo gas, roughly on the order of the Hubble time around the virial radius, results in relatively subtle changes in the outer region. As the radius decreases, this timescale decreases to 1Gyr at around 0.2 $r_{vir}$ and further reduces below 1 Gyr in the innermost region. However, our recipe that avoids over-cooling ensures that the cooling time for gas with $T> \ 5\times 10^5$ K and number density $n> 1\times 10^{-2} \rm{cm^{-3}}$ is not less than 1 Gyr. In all, the cooling time scale of the central region is shorter than the simulation run time of 4 Gyr. This is why the temperature dips and the number density is enhanced in the inner region, as shown Figure \ref{fig:without-inflow}. Meanwhile, the gas in the outer region slowly moves inward under the gravitational field as the temperature drops in the inner region. Consequently, the temperature (density) of gas in the outer region gradually increases (decreases). Moreover, higher metallicity leads to more significant changes because of more efficient cooling. However, it is important to note that at t=4.0 Gyr, the average temperature of the gas remains above $\sim 10^{6}$ K throughout, from the halo center to the region beyond the virial radius. The analysis based on the four test simulations affirms that if any cold gas appears within the halo's central region in the simulations incorporating injected cold streams, cold streams should be the game changer.

\subsubsection{Gas phases}

As the cold gas descends and interacts with the hot halo gas, undergoing processes such as compression, radiative cooling, and turbulent mixing, the gaseous medium manifests into multi-phases. To describe the evolution of different phases in the following context, we classify gaseous mediums into three different categories according to their temperature: 
\begin{enumerate}
    \item  \textit{Cold and cool}: gas with a temperature T < $10^5$ K.
    \item  \textit{Warm}: gas with $10^5 \leq$ T < $ 10^6$ K.
    \item  \textit{Hot}: gas with T $\geq 10^6$ K. 
\end{enumerate}

\section{Results}

In this section, we present our simulation results starting from the ones with a gas metallicity $Z=0.1Z_{\odot}$, where the cooling of gas in the cold stream is more efficient and hence has a higher chance to survive in the hot gaseous halos. Then, we inspect the impact of gas metallicity using other simulations.

\subsection{Overall evolution}

\begin{figure*}
    \centering
    \includegraphics[trim ={0.cm 1.0cm 0.cm 0cm},clip,width=2.0\columnwidth]{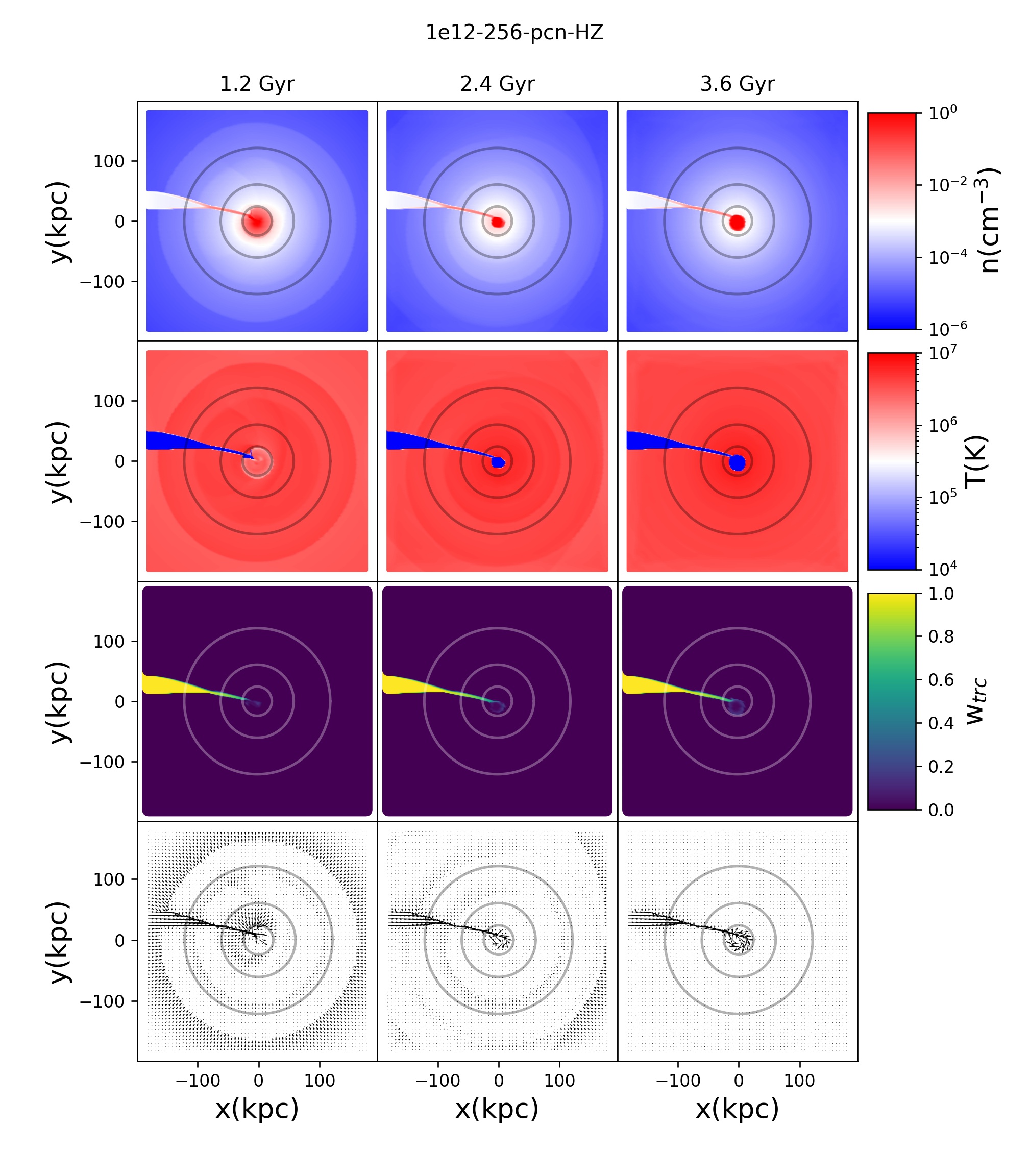}
    \caption{Slices through the central xy plane along z direction of the simulation 1e12-256-pcn-HZ, i.e., with $M_h = 10^{12} \rm{M_{\odot}}$, inhomogeneous cold streams, cooling and $Z=0.1Z_{\odot}$, but without conduction. The first and second rows show the number density and temperature, respectively. The third row is the tracer weight $w_{trc}$ (mentioned in Section \ref{sec:cold streams}), and the last row is the velocity field. The three columns from left to right show results at time 1.2Gyr, 2.4Gyr, and 3.6Gyr, respectively.}
    \label{fig:m10ma10_slice}
\end{figure*}

\begin{figure*}
    \centering  
    \includegraphics[trim ={0.cm 1.0cm 0.cm 0cm},clip,width=2.0\columnwidth]{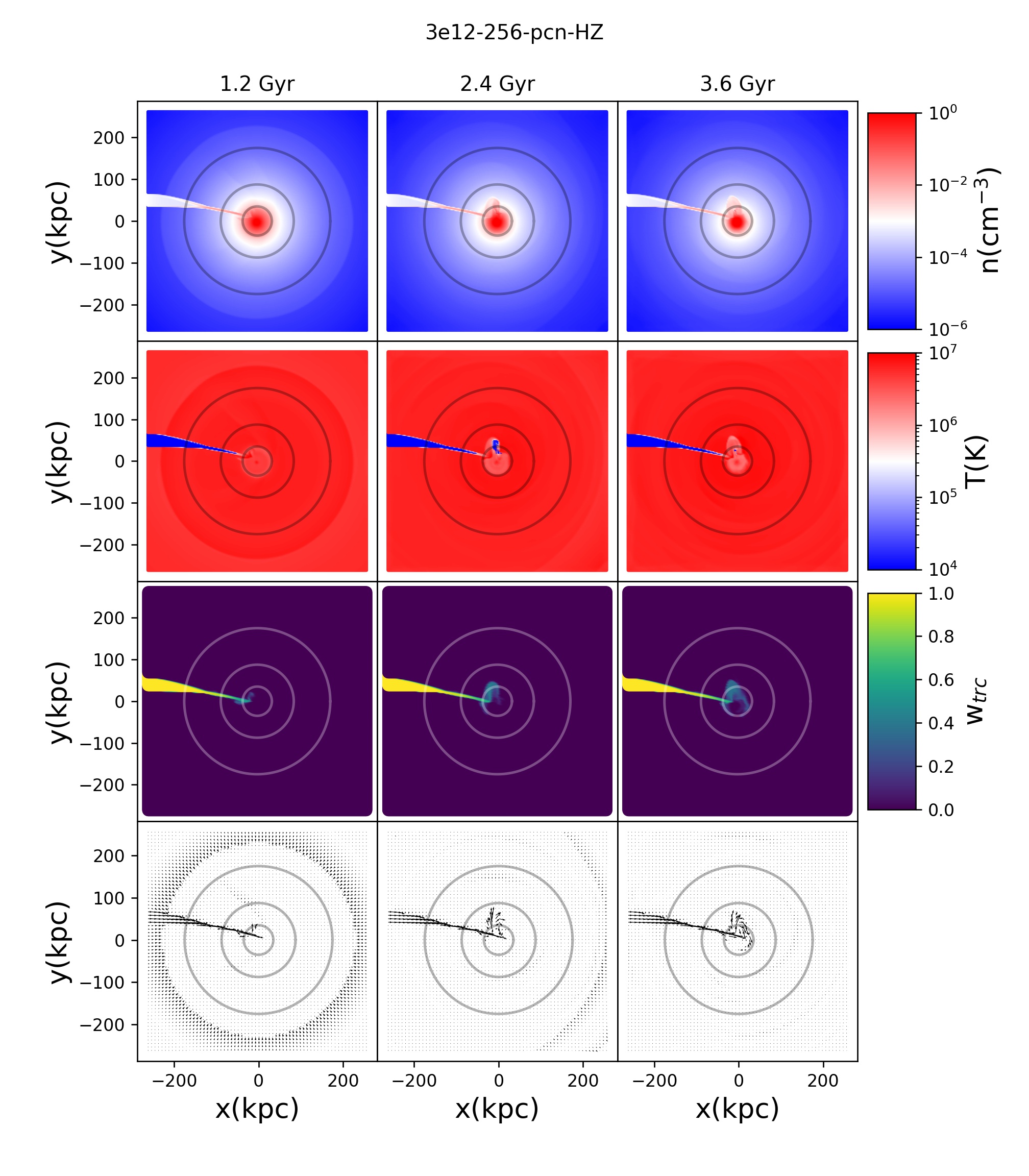}
    \caption{The same as Figure \ref{fig:m10ma10_slice} but for the simulation 3e12-256-pcn-HZ, in which $M_h = 3 \times 10^{12} \rm{M_{\odot}}$.}
    \label{fig:m10ma30-slice}
\end{figure*}

We first inspect the overall evolution of inhomogeneous cold streams in hot gaseous halos. Figure \ref{fig:m10ma10_slice} and Figure \ref{fig:m10ma30-slice} show the results with halo mass $1 \times 10^{12} \rm{M_{\odot}}$ and $3 \times 10^{12} \rm{M_{\odot}}$ respectively, adopting a metallicity of $Z=0.1Z_{\odot}$. The first and second rows in each figure show the density and temperature of the gas in a slice of depth around 1 kpc. The third and fourth rows illustrate the tracer field of gas initially injected at the boundary and the velocity fields, respectively. 

In the first $\sim 1.2$ Gyr, cold streams fell toward the central region due to gravitation in both simulations. At $t \sim 1.2 $ Gyr, a tiny fraction of cold gas appears in the central region, i.e., $r< 0.2\, r_{vir} (\sim 24\, \rm{kpc})$, of the halo with $M_h = 10^{12} \rm{M_{\odot}}$. Figure \ref{fig:m10ma10_slice} only shows the evolution of the gas in a slice. The time since the existence of cold gas in the central region may be earlier in other slices. Starting from $t=2.4 $ Gyr, cold gas accumulates gradually and occupies a considerable fraction of the volume in the central region, i.e., $r< 0.2\, r_{vir}$, of halo $M_h = 10^{12}\, \rm{M_{\odot}}$. Note that we neglected the central galaxies in our simplified model, which would somehow limit the application of our simulation in the very central region (for instance, the inner 10-20 kpc). 

The evolution of the tracer field indicates that a small fraction of gas injected at the boundary had reached the central region no later than $t=1.2$ Gyr, which was synchronous with the emergence of cold gas in the central region. The velocity field shows that the cold stream has a minor impact on the overall gas distribution, mainly limited to the stream itself and the central region after $t \sim 2.4 $ Gyr. In the central region ($r< 0.2\, r_{vir}$), the cold gas has a notable rotational velocity, which should result from the initial angular momentum.

Figure \ref{fig:m10ma30-slice} shows the gas evolution in a more massive halo with $M_h = 3 \times 10^{12} \rm{M_{\odot}}$. On the contrary, the central region remains warm and hot till the end of the simulation. Although there is some residual cold gas around 0.2 $r_{vir}$ after $t=2.4$ Gyr, the cold gas is quickly heated again. Meanwhile, the tracer field suggests that some of the gas initially injected at the boundary had also arrived at the central region of $r\lesssim 0.2\, r_{vir}$ since $t=1.2$ Gyr. However, their presence in the central region of the halo does not lead to the accumulation of cold gas therein. On the other hand, the cold stream introduces moderate changes in the gas velocity, mainly around $r \sim 0.2\, r_{vir}$. 

\begin{figure*}
    \centering  \includegraphics[trim ={0.cm 2.0cm 0.cm 0cm},clip,width=2.0\columnwidth]{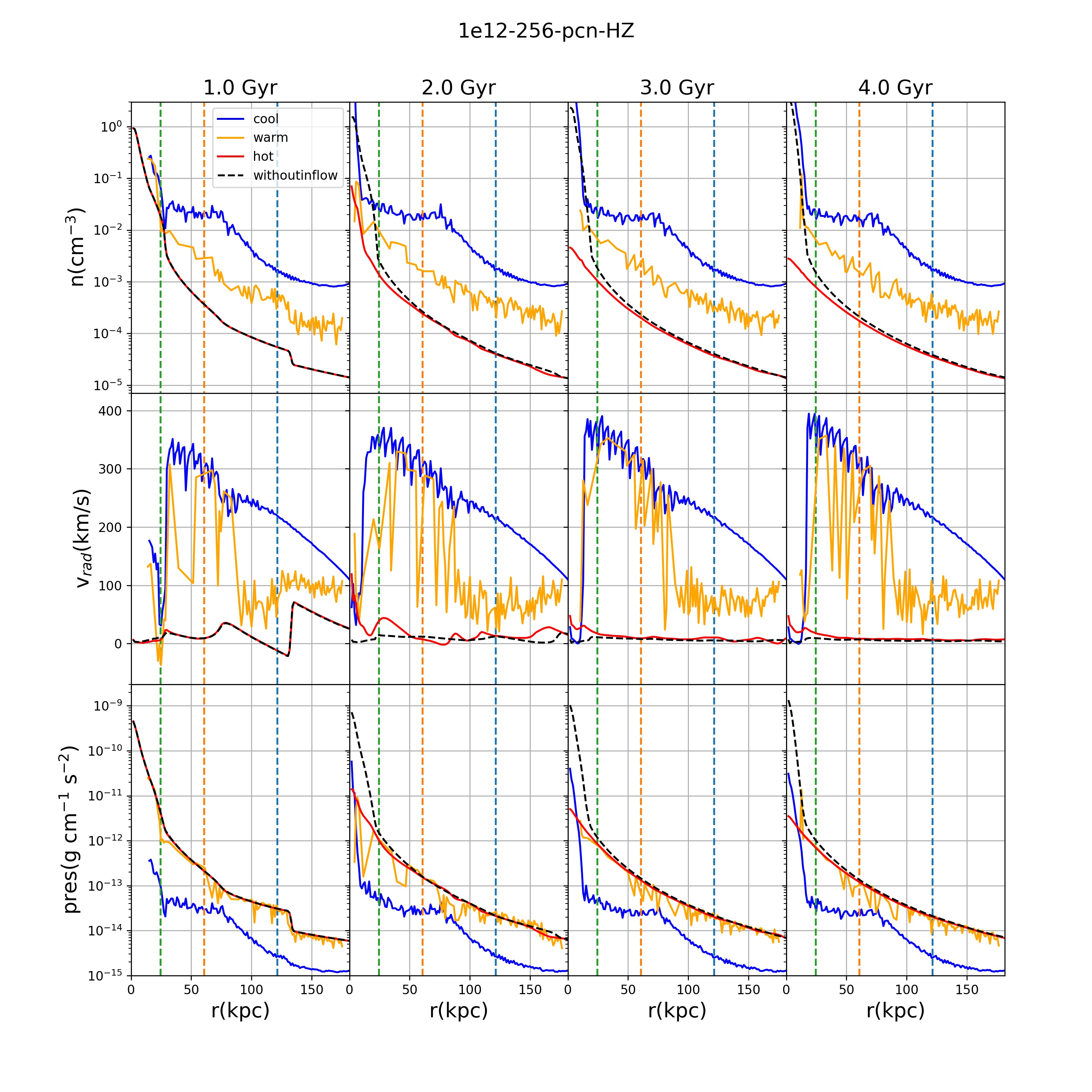}
    \caption{Rows from top to bottom show the averaged number density, radial velocity, and pressure as functions of distance from the halo center, respectively, for the simulation 1e12-256-pcn-HZ. The results at times 1.0, 2.0, 3.0, and 4.0 Gyr are presented from the left column to the right column, respectively. Red, orange and blue lines indicate hot (T$\geq 10^{6}$ K), warm ($10^{5}  \leq T < 10^{6}$ K), and cold and cool gas ($ T < 10^{5}$ K). Vertical dotted lines indicate r= 0.2 (green), 0.5 (orange), 1.0 (blue) $r_{vir}$. The black dotted line indicates the simulation results without inflow cold stream.}
    \label{fig:m10ma10-Curve}
\end{figure*}

\begin{figure*}
    \centering  \includegraphics[trim ={0.cm 2.0cm 0.cm 0cm},clip,width=2.0\columnwidth]{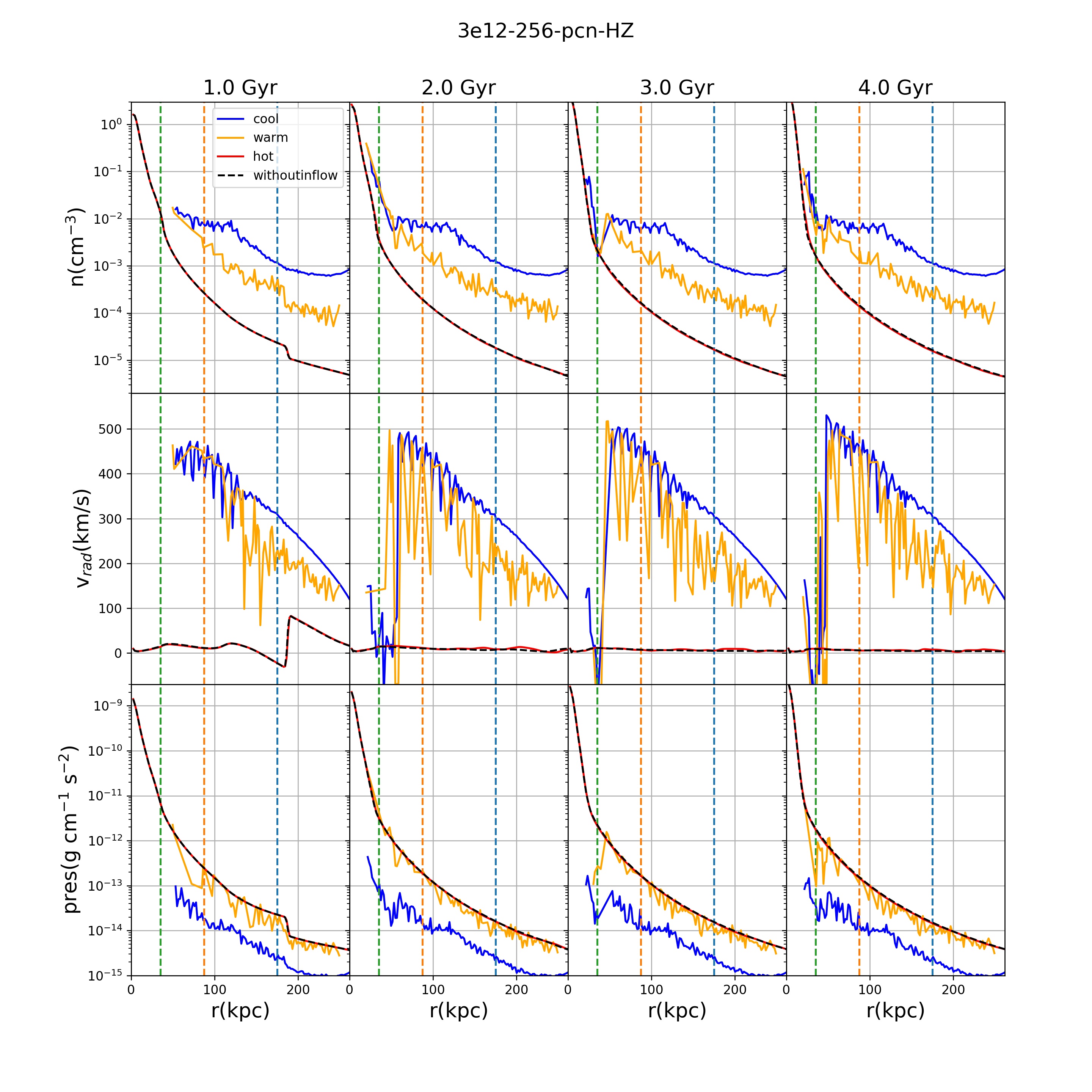}
    \caption{The same as Figure \ref{fig:m10ma10-Curve}, but for the simulation 3e12-256-pcn-HZ, in which $M_h = 3 \times 10^{12} \rm{M_{\odot}}$.}
    \label{fig:m10ma30-Curve}
\end{figure*}

In both simulations, the features of KHI were not observed. The radius of the simulated cold stream $R_{s,sim}$ can be roughly estimated by the fraction of the volume occupied by the cold gas in a particular shell. We find that the radii of the cold stream in our simulation, $R_{s,sim}$, are usually significantly greater than the critical radius $R_{s,crit}$. For instance, in the 1e12-256-pcn-HZ simulation, $R_{s,sim} \sim 40$ and 500 $R_{s,crit}$ at $r=1.0$ and 0.5 $r_{vir}$, respectively. Therefore, the role of KHI should have been suppressed by radiative cooling, which is consistent with the estimates in Section 2.2. However, it is worth noting that the limited resolution in our simulations could, to some extent, impede the growth of KHI. Furthermore, the development of KHI might have been attenuated due to the high Mach number and the considerable density contrast.

To track the evolution of cold streams, we have measured the averaged number density, radial velocity, and pressure of gas in three thermal phases as functions of distance from the halo center. The radial profiles of gas properties in halos with $M_h = 1 \times 10^{12}\, \rm{M_{\odot}}$ and $3 \times 10^{12}\, \rm{M_{\odot}}$ are shown in Figure \ref{fig:m10ma10-Curve} and \ref{fig:m10ma30-Curve} respectively. 

The density profiles in Figure \ref{fig:m10ma10-Curve} indicate that cold and cool gas appears in the central region ($r<0.2 r_{vir}$) of the halo $M_h = 1 \times 10^{12}\, \rm{M_{\odot}}$ since $t \sim 1.0 \, $ Gyr, and accumulates gradually after that. During their path toward the center of the halo, the inward radial velocity of the cold gas increases from around 200 km/s at $r\sim r_{vir}$ to $\sim 400$ km/s at $r \sim 0.1\, r_{vir}$, and then decreases sharply to $\sim 0$ km/s at a smaller radius. The acceleration and deceleration are driven by the gravitation and pressure gradient, respectively. Given the temperature and velocity of gas, the cold and cool gas is falling inward with a supersonic speed (the Mach number of the gas in the cold stream $M_s = v_s/c_s$ varies from 12.3 to 24.6, and the background Mach number $M_b=v_s/c_h$ varies from 1.23 to 2.47) in the range $ 0.1 r_{vir}<r <r_{vir}$. Meanwhile, as can be seen from the velocity and pressure plots in the second and third rows, a tiny fraction of the gas had turned to the warm phase, likely from the cold and hot phases. On average, 10\% of the hot gas turns into the warm phase. The warm gas moves inward at a speed moderately slower than that of the cold gas while it is in thermal equilibrium with the hot phase.

Figure \ref{fig:m10ma30-Curve} indicates a different history for the halo with $M_h = 3 \times 10^{12}\, \rm{M_{\odot}}$. Cold and warm gases can not exist in the central 20 kpc, i.e., 0.1 $r_{vir}$, till the end of the simulation, despite their radial velocities in the outer region being faster than in $M_h = 10^{12}\, \rm{M_{\odot}}$. Therefore, the cold and warm gas at r>20 kpc should have been heated to the `hot' phase. As we have discussed in the previous context, the third row in Figure \ref{fig:m10ma30-slice} demonstrates that some of the gas that was initially injected at the boundary as the cold stream had arrived at the central region earlier than $t=1.2$ Gyr. However, their presence does not lead to the emergence of cold and cool phases in the central region of r<20 kpc, i.e., $r\lesssim 0.1 r_{vir}$. 

The bottom row of Figure \ref{fig:m10ma10-Curve} and \ref{fig:m10ma30-Curve} indicate that the thermal pressure of the cold stream is lower than that of the hot halo gas. This discrepancy can be traced back to the initial conditions for the cold streams and the hot halo at the boundary. The parameters settings for the hot halo are detailed in \ref{sec:hot-halo}, and the cold streams are listed in \ref{sec:cold streams}. The adopted properties of the cold streams and hot halo gas are similar to previous cosmological hydrodynamical simulations. It is important to note that the thermal pressures of the two phases are not expected to be identical, given their distinct motions. The hot phase remains nearly static, while the cold phase descends inward with a velocity exceeding 100 km/s. In addition, the imbalance in thermal pressure between two phases should have shaped the cone-like geometry of a cold stream, as illustrated in previous works (e.g., \citealt{2009Natur.457..451D, 2013MNRAS.429.3353N}). 

\subsection{Competition between cooling and compressional heating}

\begin{figure*} 
    \centering
    \includegraphics[trim ={1cm 4cm 1cm 0cm},clip,width=2.0\columnwidth]{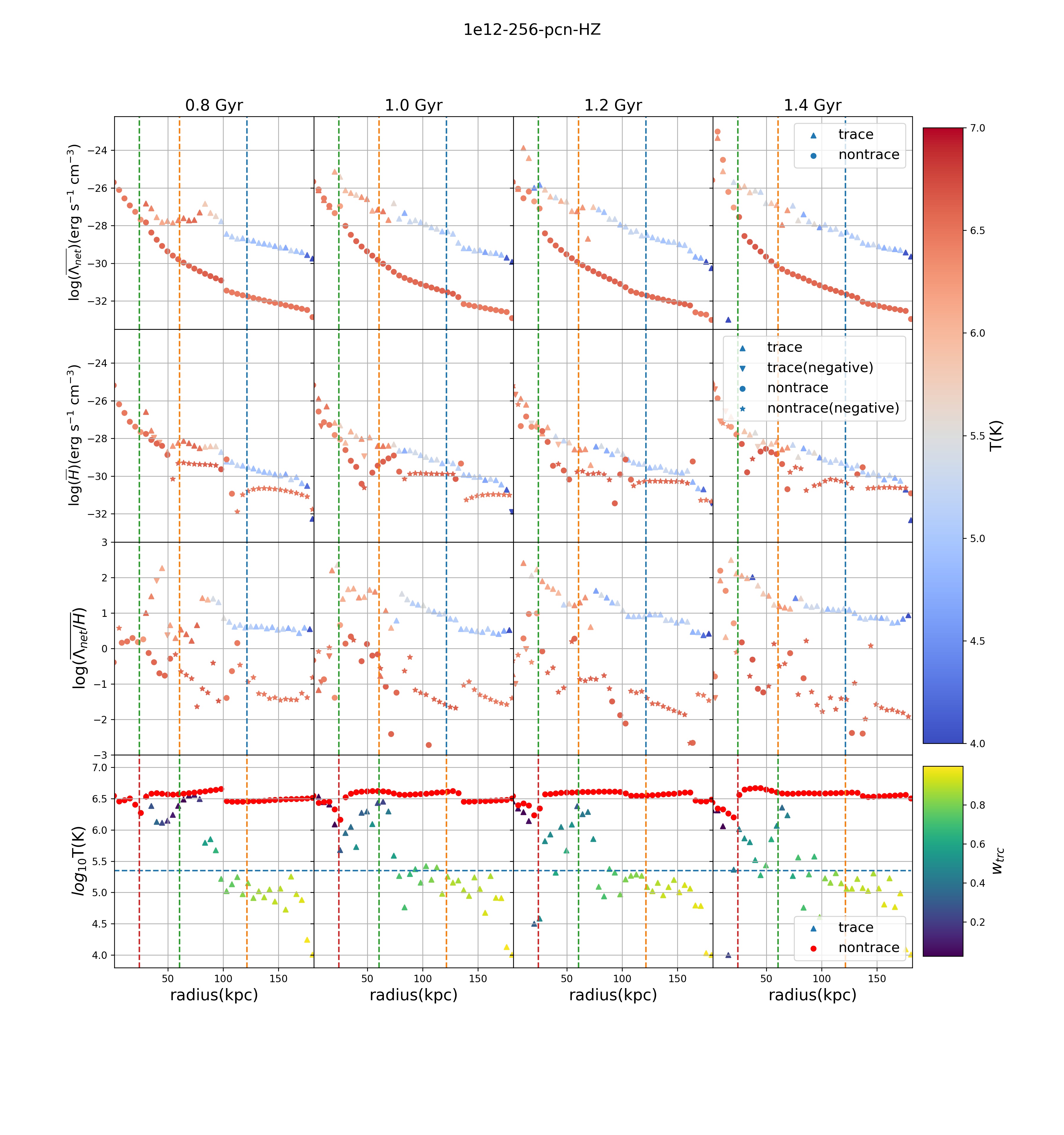}
    \caption{The first, second, third, and fourth row show the mean cooling, compression rate, ratio of cooling/compression, and the average temperature as functions of the radius from halo center, respectively, at time 0.8, 1.0, 1.2, 1.4 Gyr (from left to right) for the simulation 1e12-256-pcn-HZ, in which $M_h = 10^{12} M_{\odot}$. The data points in the first three rows are color-coded by the average temperature. The triangles (circles) represent the average values of gas cells (do not) containing tracers of injected cold streams. In the second row, the downward triangles (stars) indicate that the gas cells (do not) contain a tracer of injected cold streams and have a negative compression rate. Vertical dotted lines indicate the radius of r= 0.2 (green), 0.5 (orange), 1.0 (blue) $r_{vir}$.} 
    \label{fig:m10ma10-lamb}
\end{figure*}

\begin{figure*}
    \centering
    \includegraphics[trim ={1cm 4cm 1cm 0cm},clip,width=2.0\columnwidth]{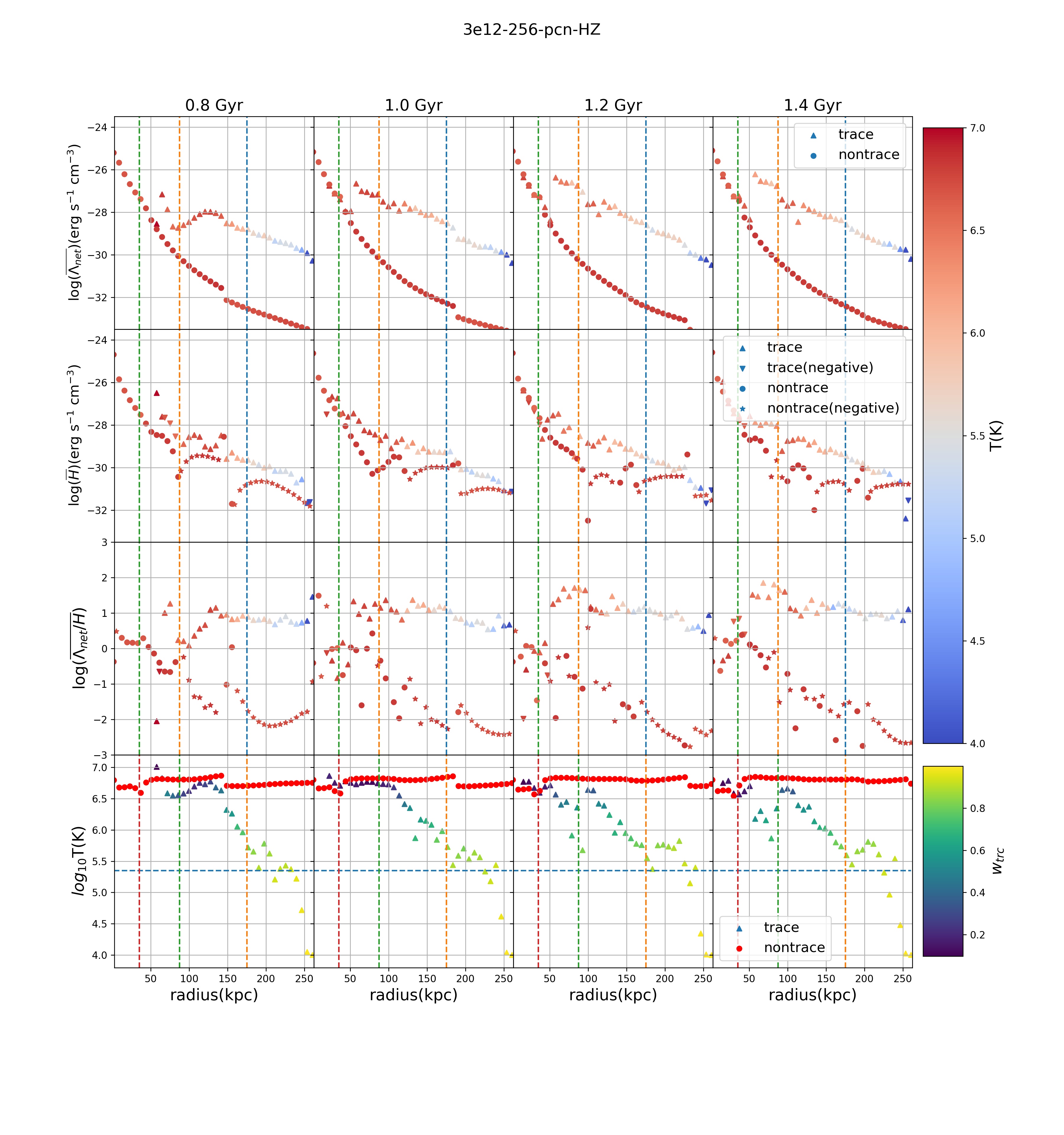}
    \caption{The same as Figure \ref{fig:m10ma10-lamb}, but for the simulation 3e12-256-pcn-HZ, in which $M_h = 3 \times 10^{12} M_{\odot}$. }
    \label{fig:m10ma30-lamb}
\end{figure*}

In the case of spherical accretion, \citet{2003MNRAS.345..349B} had demonstrated that the rivalry between radiative cooling $\Lambda_{net}$ and the compress of the gas determines the stability of the accretion shock. According to the estimation on the typical time scales of multi-physical processes given in Section 2.2, the competition between radiative cooling $\Lambda_{net}$ and compressional heating caused by gas pressure should be the key to determining the thermal evolution of gas in the cold stream. The compressional heating rate can be given by (\citealt{2003MNRAS.345..349B}) 
\begin{equation}
    \Bar{H} = -p\triangledown V\,.
\end{equation}

Before we present the detailed results on cooling and compression heating, we should remind the readers again that this work focusses on the evolution of cold streams in the CGM environment. In this work, the model in the central 0.2 $r_{vir}$ region is oversimplified. If there are other heating sources in the central region, such as feedback from star formation and supermassive black holes, they should be added to the RHS of the above heating equation. The impact of central feedback on the cold stream is beyond the scope of this work. 

We conduct a quantitative study on the gas cooling and compressional heating rates in our simulations. Moreover, we divide the gas into two categories, one with tracer density greater than 0, i.e., $\omega_{trac}>0$, and the other is tracer-free. The former category contains gas injected at the boundary, while the latter is the origin of hot halo gas that has not yet mixed with the injected gas. From the snapshots of the simulation, we use post-simulation analysis to measure the mean cooling rate, compressional heating rate, the ratio of cooling/compression, and the average temperature of the two categories of gas as functions of the radius from the halo centre. The corresponding values are volume-weighted means that take gas cells within different radius bins, with an interval of $\delta r=0.01\,r_{vir}$. The results of the run with halo mass $M_h = 1\times 10^{12} \rm{M_{\odot}}$ and gas metallicity $z=0.1Z_{\odot}$ are shown in Figure \ref{fig:m10ma10-lamb}. From the left to the right columns, the results at times 0.8, 1.0, 1.2, and 1.4 Gyr are illustrated by turns. The lines in the first three rows are colour-coded with the average temperature. The solid triangle (circle) indicates the gas with $\omega_{trac}>0$ ($\omega_{trac}=0$). Because not all gas cells experience compression, there are some negative values for the compressional heating rate and the cooling/compression ratio. We mark these data points with downward triangles and pentagrams, showing their absolute values.

In the first 0.8 Gyr, the gas in the cold stream is still in its path, falling toward the centre. At the same radius, the mean cooling rate of gas with tracer density $\omega_{trac}>0$ is larger than their counterparts with $\omega_{trac}=0$ by about 2-3 orders of magnitude. The probable reason is as follows. The gas with $\omega_{trac}>0$ is the mixture of the injected cold gas and the origin hot halo gas and, therefore, has a temperature between $10^4$ K (cold stream) and $\sim 10^6$ K (hot halo). When the temperature of mixed gas approaches the plateau in the cooling curve shown in Figure \ref{fig:coolingfunction}, i.e., $10^{4.9}-10^{5.4}$ K for metallicity $Z=0.1\, Z_{\odot}$, the cooling rate would increase dramatically. The mean cooling/compression ratio of the mixed gas ranges from 1 to 100, which is also $\sim 100$ times higher with respect to the gas with $\omega_{trac}=0$ at the same radius. In the central region where the gas in the cold stream has not reached out (there is no sign of tracer at $r<0.2r_{vir}$), the mean cooling/compression ratio is around 1, indicating that the mean cooling rate is either marginally balanced with or is moderately higher than the mean compressional heating rate for the hot gas therein. This is consistent with the results of relevant timescales presented in Section 2.2. 

At $t \sim 1.0 $ Gyr, the frontier of the cold stream has reached the very central region of the halo. Since then, the cooling of the mixed gas in the central region has been boosted and its rate has grown considerably higher than the compressional heating rate. The corresponding cooling/compression ratio can be as large as 100 at $t \gtrsim 1.0 $ Gyr. This remarkable feature mainly results from the temperature of the mixed gas in the central region becoming lower than $10^6$ K and climbing to the peak of the cooling curve. Later, the mixed gas within $r < 20 $ kpc experiences runaway cooling at a cooling/compression ratio greater than 100 since the temperature of some mixed gas is close to the plateau of cooling rate, i.e., $10^{4.9-5.4}$ K. This result also agrees with the analysis of cooling and compressional heating timescales introduced in Section 2.2. Consequently, the mean temperature of mixed gas can be lower than $10^5$ K in the central region since $t = 1.4$ Gyr.

The cooling and compressional heating rates of the gas in the halo $M_h= 3\times 10^{12} \rm{M_{\odot}}$ are shown in Figure \ref{fig:m10ma30-lamb}. In the first 1.0 Gyr, the cooling efficiency of gas is similar to that of the halo $M_h= 10^{12} \rm{M_{\odot}}$. At $t \sim 1.0 $ Gyr, cold gas injected at the boundary had arrived at the inner region of the halo, as indicated by the tracer field. In contrast to the case in the halo of $M_h= 10^{12} \rm{M_{\odot}}$, the increment in the cooling rate of mixed gas was moderate and comparable with the rise of compressional heating rate. Therefore, the cooling/compression ratio of the mixed gas fluctuates around 1.0, which is lower than that in the outside region. Consequently, the gas in the central region of the halo remains warm and hot till the end of the simulation. The difference between the two halos mentioned above is that a deeper potential well hosts a gaseous halo with higher temperatures, leading to a larger falling velocity and stronger compressional heating. Thus, the temperature of the mixed gas is far away from the plateau in the cooling curve for a halo with $M_h= 3\times 10^{12} \rm{M_{\odot}}$, resulting in a significantly longer cooling time. 

\subsection{Supply of cold and cool gas}

\begin{figure*}
    \centering
    \includegraphics[trim = {3.2cm 2cm 4cm 0cm},clip,width=2.0\columnwidth]{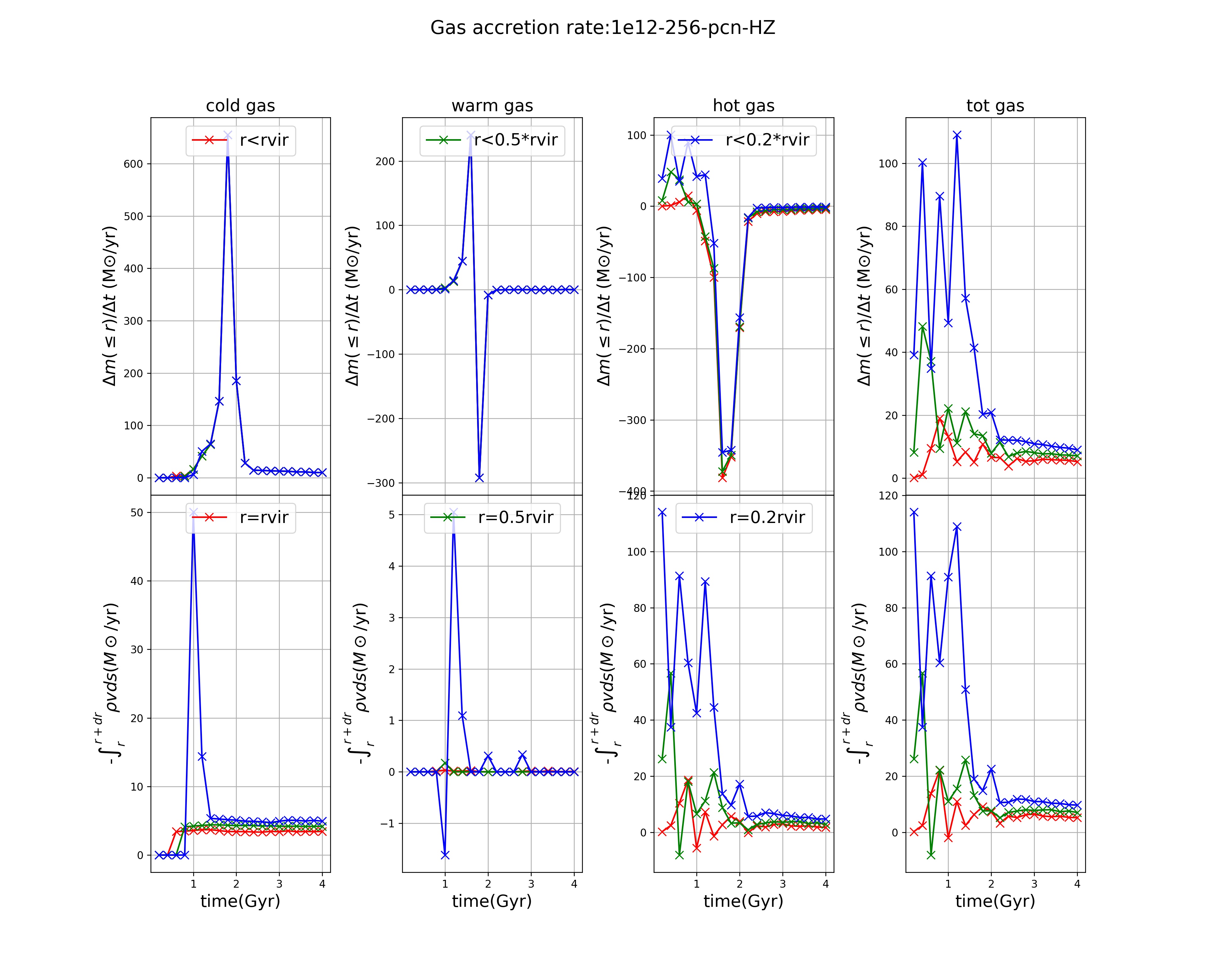}
    \caption{Gas accretion rate for the simulation 1e12-256-pcn-HZ, in which $M_h=10^{12} M_{\odot}$. The top row indicates $\bar{\dot{M}}(r,t)$ within r=1.0$r_{vir}$(red), r=0.5$r_{vir}$(green) and r=0.2$r_{vir}$(blue). The bottom row presents the instant accretion rate $\dot{M}(r,t)$ at r=1.0$r_{vir}$, r=0.5$r_{vir}$ and r=0.2$r_{vir}$. See equation \ref{eqn:ave_accretion} and \ref{eqn:ins_accretion} for the definition of accretion rate. From the left column to the right, the plots show the cold and cool, warm, hot, and all the gas, respectively. }
    \label{fig:m10ma10-accre}
\end{figure*}

\begin{figure*}
    \centering
    \includegraphics[trim = {3.2cm 2cm 4cm 0cm},clip,width=2.0\columnwidth]{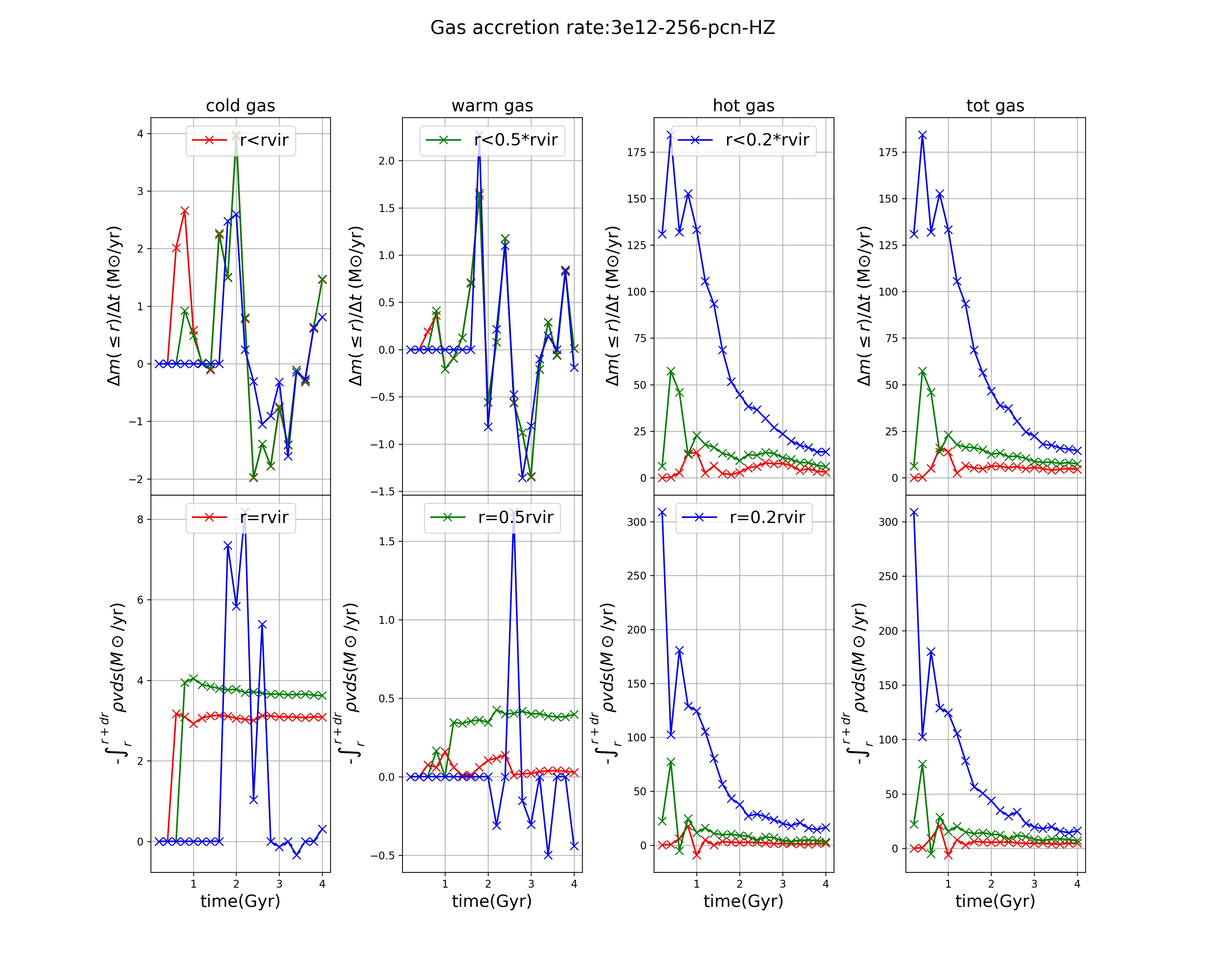}
    \caption{The same as Figure \ref{fig:m10ma10-accre}, but for the simulation 3e12-256-pcn-HZ, in which $M_h=3 \times 10^{12}\, \rm{M_{\odot}}$. }
    \label{fig:m10ma30-accre}
\end{figure*}

\begin{figure*}
    \centering
    \includegraphics[trim = {3cm 2cm 4cm 3cm},clip,width=2.0\columnwidth]{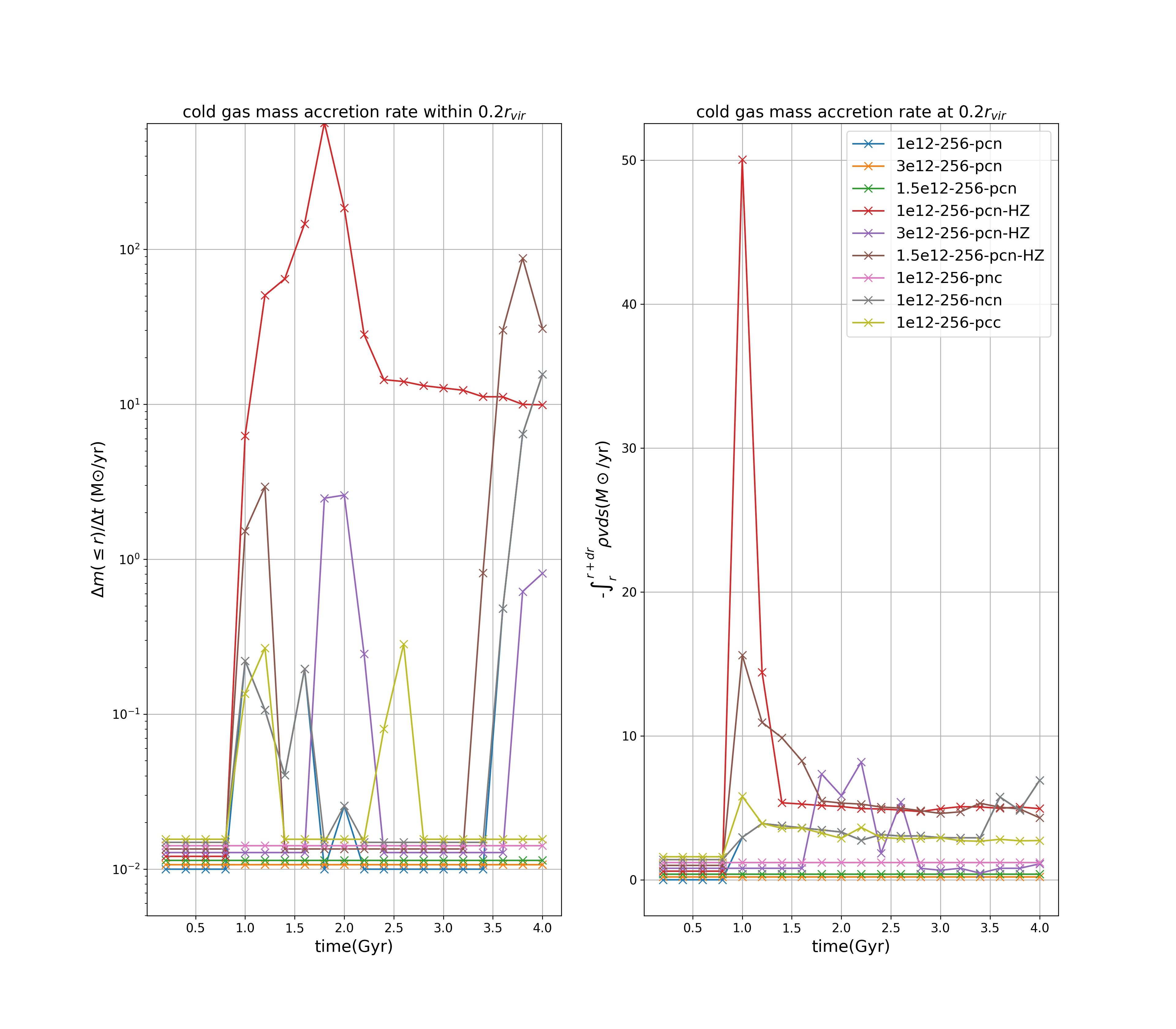}
    \caption{The accretion rate of cold and cool gas at 0.2 $r_{vir}$ measured in nine simulations. The setting of different simulations can be found in Table \ref{tab: Simulations Groups} and Section 2.2. The left and right panels present $\bar{\dot{M}}(r<0.2r_{vir},t)$ and $\dot{M}(r=0.2r_{vir},t)$ respectively. See Equations \ref{eqn:ave_accretion} and \ref{eqn:ins_accretion} for definition. Note that a lower limit of 0.01 is set for dm/dt to enable full coverage in the left panel. Meanwhile, to prevent overlap, the lines have been slightly shifted upward in both panels. The magnitude of the shift for each line can be determined by the results at $t<0.5$ Gyr, where the actual value is 0 for all simulations. The flat lines in both panels indicate no cold gas flow into or within the central 0.2 $r_{vir}$ region. A value exceeding the value at $t<0.5$ Gyr signifies cold gas inflow or generation of cold gas through cooling within the central 0.2 $r_{vir}$ region at that particular time.}
    \label{fig:cold and cool gas accretion rate at 0.2rvir}
\end{figure*}

Cold streams play an essential role in galaxy formation and evolution because they serve as the channel to provide cold and cool gas for the central galaxy in halos with a mass below and moderately higher than $10^{12} \rm{M_\odot}$ at $z \gtrsim 1$. The critical factor is the mass rate of the cold and cool gas at which the central region of the halo can obtain. We probe the quantitative change of the gas mass in three phases, i.e., cold, warm, and hot, at three radii, i.e., r = 1.0, 0.5, and 0.2 $r_{vir}$ respectively. We have measured the changes in two different ways. One is the averaged change of gas mass within a certain radius between two successive snapshots and is denoted as:
\begin{equation}
\bar{\dot{M}}(r,t)=\frac{M(r,t)-M(r,t-\delta t)}{\delta t},
\label{eqn:ave_accretion}
\end{equation}
where $M(r,t)$ is the total mass of gas in a particular phase within the radius $r$ at the simulation time $t$, $\delta t$ is the time interval between snapshots. The other method describes the instant inflow rate at the spheres with three given radii, i.e., 
\begin{equation}
\dot{M}(r,t)=-\int \rho \vec{v} \cdot d\vec{S}.
\label{eqn:ins_accretion}
\end{equation}
The evolution of $\bar{\dot{M}}(r,t)$ and $\dot{M}(r,t)$ in the two simulations with gas metallicity $z=0.1 Z_{\odot}$, and halo mass $M_h=10^{12} M_{\odot}$ and $3 \times 10^{12} M_{\odot}$ are shown in Figure \ref{fig:m10ma10-accre} and \ref{fig:m10ma30-accre} respectively. In each figure, the upper row indicates $\bar{\dot{M}}(r,t)$ at r = 1.0 $r_{vir}$ (red line), r = 0.5 $r_{vir}$ (green line) and r = 0.2 $r_{vir}$ (blue line). Meanwhile, the bottom row presents the instant inflow rate $\dot{M}(r,t)$. From the left to right columns, the plots show the measurements of cold and cool, warm, hot, and all the gas, respectively. 

For the halo with $M_h=10^{12}\, \rm{M_{\odot}}$, the instant inflow rate suggests that the cold and cool gases begin to flow into $r=1.0, 0.5, $ and 0.2 $r_{vir}$ at the time $0.6, 0.8$ and $1.0$ Gyr, respectively. From $t=1.4 $ Gyr to the end of the simulation, the inflow rate of cold and cool gas remains around 3.0 $\rm{M_{\odot}/yr}$ at $r=1.0\, r_{vir}$, and $\sim 4\, \rm{M_{\odot}/yr}$ at $r=0.5\, r_{vir}$. At $r = 0.2\, r_{vir}$, the accretion rate of cold and cool gas increases from 0 at $t=0.8$ Gyr to 50 $\rm{M_{\odot}/yr}$ at $t=1.0$ Gyr, and then decreases sharply to 5.0 $\rm{M_{\odot}/yr}$ at $t=1.4$ Gyr. This violent and short change is likely triggered by the arrival of the frontier of the cold stream at $r=0.2\, r_{vir}$, combining the results presented in Figures \ref{fig:m10ma10_slice}, \ref{fig:m10ma10-Curve} and \ref{fig:m10ma10-lamb}. Between $t=1.4$ and $t=4.0$ Gyr, $\dot{M}(r,t)$ of cold and cool gas at 0.2 $r_{vir}$ stabilise at $\sim 5\, \rm{M_{\odot}/yr}$. Note that the injection rate of the cold gas at the boundary is $\sim 3 M_{\odot}/yr$, which is equal to the instant accretion rate of cold and cool gas at $r=1.0\, r_{vir}$, but lower than the accretion rate at $r=0.5$ and $0.2\, r_{vir}$ by $\sim 30-50\%$. Cooling should be responsible for the increment from $r=1.0\, r_{vir}$ to $r=0.5\, r_{vir}$ and $r=0.2\, r_{vir}$.

On the other hand, the mass of cold and cool gas barely grows in the halos ($r<r_{vir}$) before $t=1.0$ Gyr, as shown by the top-left panel in Figure \ref{fig:m10ma10-accre}. From $t=1.2$ Gyr to $t=2.2$ Gyr, the total mass of cold and cool gas in the central region, i.e., $r \leq 0.2\, r_{vir}$ increases at a rate greater than 50 $\rm{M_{\odot}/yr}$. The dominant cause should be the runaway cooling of the hot gas after mixing with the cold gas, as demonstrated by the plots in the second and third columns. After $t=2.2$ Gyr, cold and cool gases accumulate steadily within $r<0.2\, r_{vir}$, at a rate approximately equal to the instant inflow rate at $r=0.2\, r_{vir}$. The mass of warm gas within three radii changes dramatically between $t=1.0$ Gyr and $t=2.0$ Gyr. The probable reason is that the warm phase is the temporary transition phase between cold and hot and cannot exist stably. During the epoch $t=1.0$ and $t=2.0$ Gyr, a considerable amount of hot gas first cools to the warm phase, resulting in a peak increment of warm gas, and then the warm gas is further cooled to the cold gas, resulting in a dip of warm gas at 1.8 Gyr. Between $t=1.0$ Gyr and $t=2.0$ Gyr, the mass of hot gas within the halo experiences a violent decline. In addition to the runaway cooling in the very central region, there are several other driving factors. Any imbalance between the gravitation and pressure gradient would lead to the inflow of hot halo gas. Furthermore, part of the gas in the cold stream was heated up to the hot phase. 

We then extended our investigation to the case where $M_h=3\times 10^{12}\, \rm{M_{\odot}}$. The bottom left panel in Figure \ref{fig:m10ma30-accre} shows that cold and cool gas flows steadily into spheres with radii of $r=1.0$ and $0.5$ $r_{vir}$ at rates of 3 $\rm{M_{\odot}/yr}$ and 4 $\rm{M_{\odot}/yr}$ from approximately $t\sim 0.4$ Gyr to the end of the simulation. However, $\dot{M}(r,t)$ of cold and cool gas at the radius of $r=0.2 r_{vir}$ is positive only in the time range from t=1.6 Gyr and t=2.8 Gyr. Subsequently, this inflow rate decreases to nearly zero after t=2.8 Gyr. It is worth noting that the accumulation of cold and cool gas within $r=0.2\, r_{vir}$ occurs at a considerably slower pace compared to the case with $M_h=10^{12} \rm{M_{\odot}}$, which is also indicated in the first row of Figure \ref{fig:m10ma30-Curve}. Furthermore, no cold and cool gas is found within $r \leq 20$ kpc at $t=4\, $ Gyr. Along their streamlines towards the centre of the halo, the gas injected from the boundary experiences a more significant heating due to compression than those in halos with $M_h=10^{12} \rm{M_{\odot}}$. Interestingly, the variation in the warm gas throughout the simulation is rather gentle. The hot gas exhibits a notable inflow into the central region ($r \leq 0.2 r_{vir}$) during the initial 2 Gyr, gradually slowing down thereafter. This inflow can probably be attributed to the imbalance between the gas pressure and the gravitational potential.

Briefly, we find that cold streams are highly effective in transporting cold gas to the central region of a halo ($r<0.2, r_{vir}$) with a mass of $M_h=10^{12} \rm{M_{\odot}}$. The delivery rate is equal to and sometimes exceeds the injection rate at $r > r_{vir}$. Moreover, the cold stream can lead to a runaway cooling of the hot gas within $0.2, r_{vir}$ of the halo with $M_h=10^{12} \rm{M_{\odot}}$. However, in the case of a halo with a larger mass, $M_h=3\times 10^{12} \rm{M_{\odot}}$, the cold stream is substantially disrupted by the heating due to compression, occurring at a radius slightly larger than 0.2 $r_{vir}$. Consequently, it fails to supply cold and cool gas into the central region of the halo. 

\subsection{The transition halo mass and impact of metallicity}

The results presented in the last sections show that the fate of cold streams in our simulations is mainly determined by the rivalry of cooling and compression of falling gas. The latter is governed by the halo mass, whereas the former is closely related to the metallicity of gas. This result is generally consistent with the bimodal accretion scenario proposed in previous theoretical and cosmological simulation studies (e.g., \citealt{2003MNRAS.345..349B}), where spherical accretion was studied. Previous studies have concluded that there is a maximum halo mass, $M_{stream}$,  below which a cold stream can exist. To explore the transition halos mass that the cold stream no longer survives in the central region of halos for the metallicity Z=0.1 $Z_{\odot}$, we have also performed a simulation with the same metallicity but for a halo mass $1.5\times 10^{12}\, \rm{M_{\odot}}$. For the sake of clarity, the plots of the slices and profiles are not shown. Figure \ref{fig:cold and cool gas accretion rate at 0.2rvir} shows the accretion rate of cold and cool gas at 0.2 $r_{vir}$ for different simulations. Note that, to avoid visual overlap, the lines have been slightly shifted upward in both panels of Figure \ref{fig:cold and cool gas accretion rate at 0.2rvir}. The magnitude of the offset for each line can be determined by the results at $t<0.5$ Gyr, where the actual value is 0 for all simulations. In four simulations, there is no cold gas supply to the central region until the end. In the other five simulations, the cold gas accretion rate shows significant fluctuations. In addition, the limited number of stored snapshots have caused occasional spikes. For the simulation with $M_h = 1.5\times 10^{12} \, \rm{M_{\odot}}$ and Z=0.1 $Z_{\odot}$ (1.5e12-256-pcn-Hz), the instant accretion rate remains 0 till $t\approx 0.8$ Gyr, then increases rapidly to $\sim 15 \rm{M_{\odot}/yr} $ at $t\approx 1.0$ Gyr, but then decreases to $\sim 5 \rm{M_{\odot}/yr} $ afterward. However, the total mass of cold and cool gas in the central region only gains between 0.8-1.4 Gyr and after 3.4 Gyr. That is, the cold gas supply to the central region is not continuous. Therefore, we speculate that the transition mass $M_{stream}$ is very likely close to $1.5\times 10^{12}\, \rm{M_{\odot}}$ for Z=0.1 $Z_{\odot}$. 

The simulations introduced in the previous sections adopt a gas metallicity Z=0.1 $Z_{\odot}$. To investigate the effect of metallicity, we have performed another set of simulations for halos with masses of 1.0, 1.5, and $3.0 \times 10^{12}\, \rm{M_{\odot}}$, but with a lower metallicity of Z=0.001 $Z_{\odot}$. Plots of radial profiles, cooling and compression, and accretion rates for the halo mass 1.0 $\times 10^{12}\, \rm{M_{\odot}}$ can be found in the appendix \ref{sec:appdenix-pcn}. The accretion rate of cold and cool at 0.2 $r_{vir}$ for these three simulations is also shown in Figure \ref{fig:cold and cool gas accretion rate at 0.2rvir}. For a halo mass $1.5 \times 10^{12}\, \rm{M_{\odot}}$, the cold stream cannot bring cold and cool gas into the central region with Z=0.001 $Z_{\odot}$. Nevertheless, a cold stream can deliver cold and cool gas into the central region of the halo with $M_h = 1.0 \times 10^{12} \, \rm{M_{\odot}}$, although the accumulation of cold gas only starts at $t\sim 3.4$ Gyr. In contrast to the high-metallicity case, runaway cooling is not observed. We conclude that the transition halo mass $M_{stream}$ for Z=0.001 $Z_{\odot}$ is between $1.0\times 10^{12}\, \rm{M_{\odot}}$ and $1.5\times 10^{12} \, \rm{M_{\odot}}$, somewhat lower than the case with Z=0.1 $Z_{\odot}$. 

Obviously, gas metallicity also plays a nonnegligible role in the evolution of cold streams. Higher metallicity results in a higher cooling rate and shorter cooling timescale, as indicated in Table \ref{tab: timescale}. The strength of compressional heating remains constant for a given halo mass, but the gas will cool more rapidly with increased metallicity. Consequently, the cold stream can penetrate the hot gaseous halo, causing the transition mass to rise with the gas metallicity. Our study assumes a uniform metallicity for the injected cold gas and the surrounding hot gaseous halo. However, it is important to note that the IGM, CGM, and ISM can exhibit varying metallicities. For a more refined modeling approach, it may be necessary to consider gas metallicity as a variable and track its evolution.

The transition mass found here for both metallicities is in general in agreement with the theoretical expectation (e.g., \citealt{2006MNRAS.368....2D}) and recent results obtained by \citet{2022A&A...661L...7D, 2022ApJ...926L..21D}, i.e. $2\times 10^{12}\,\rm{M_{\odot}}$.

\section{Discussion}
In addition to the simulations presented in the last section, several simulations have been carried out to explore the effects of thermal conduction and resolution. We discuss the impacts of these factors and the limitations of our study in this section.  

\subsection{Conduction}

\begin{figure*}
    \centering  \includegraphics[trim ={0.cm 2.0cm 0.cm 0cm},clip,width=2.0\columnwidth]{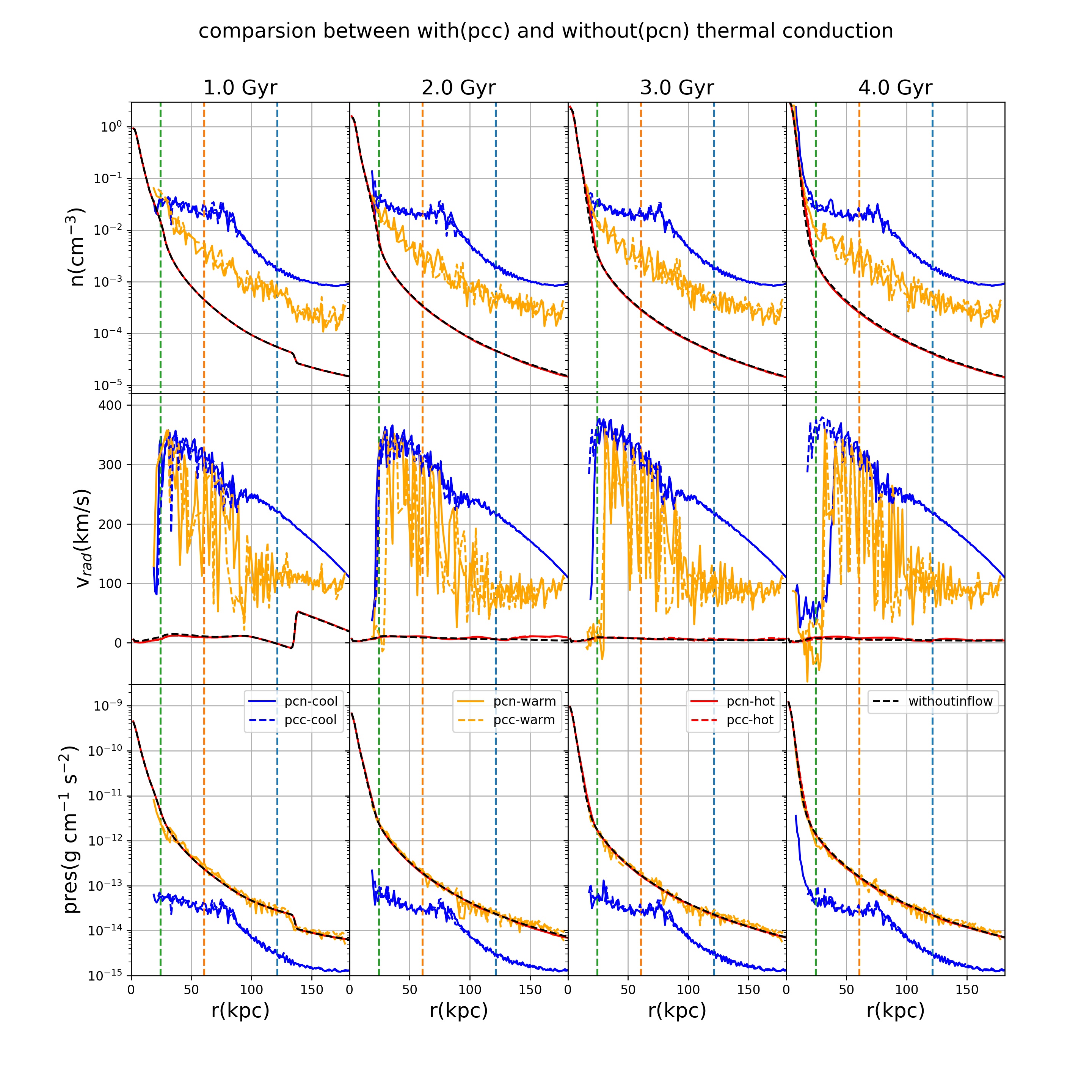}
    \caption{Rows from top to bottom show the averaged number density, radial velocity, and pressure of three gas phases as functions of distance from the halo center, respectively, for the simulation 1e12-256-pcn. The results at times 1.0, 2.0, 3.0, and 4.0Gyr are presented from the left column to the right column, respectively. The red, orange, and blue lines indicate hot (T$\geq 10^{6}$ K), warm ($10^{5}  \leq T < 10^{6}$ K), and cold and cool gas ($ T < 10^{5}$ K). Solid (dotted) lines represent simulations with (without) thermal conduction. The vertical dotted lines indicate a radius of r= 0.2 (green), 0.5 (orange), 1.0 (blue) $r_{vir}$ respectively. The black dotted line indicates the simulation results without inflow cold stream.}
    \label{fig:pcn-pcc-Curve}
\end{figure*}

Thermal conduction has been found to play an important role in the evolution of cold gas clumps within a hot gas environment (e.g., \citealt{Armillotta_2017}). Thermal conduction may impede the development of hydrodynamical instabilities at the interface of the stream and CGM (\citealt{ledos2023stability}), keeping the stream compact and therefore more difficult to destroy and thus helping the streams penetrate the central region of halo in the cold phase. One of our simulations, `1e12-256-pcc', has turned the thermal conduction on, in contrast to the simulation `1e12-256-pcn'. Figure \ref{fig:pcn-pcc-Curve} compares these two simulations on the evolution of the mean density, radial velocity, and pressure of three gas phases. Thermal conduction enables the stream to flow deeper into the central region of halo (reach out r$=0.1r_{vir}$) with a somewhat higher speed at the end of the simulation. But, as we can see, including thermal conduction has induced a minor impact on the overall evolution of cold streams. In addition, a slight change can be found in the supply of cold gas to the central region, as shown in Figure \ref{fig:cold and cool gas accretion rate at 0.2rvir}. The main reason is that cooling is more efficient than thermal conduction, as manifested by the corresponding time scales shown in Section 2.2. 

The role of thermal conduction may probably have been underestimated as a result of the limited spatial resolution. In \cite{Armillotta_2017}, the resolution could reach scales of around one pc. As mentioned in \ref{sec:resolution}, a resolution of $\Delta x$ < 100 pc is needed to fully resolve the small-scale properties of cold gas. In contrast, the resolution in this work is only around one kpc. Therefore, fine structures on scales smaller than one kpc are inevitably smoothed. Consequently, the effect of thermal conduction is likely damped somewhat in the simulation `1e12-256-pcc', as well as the KHI. Increasing the simulation resolution to pc scale, possibly by adaptively refined grids in the future study, would provide a more accurate estimation of the effects of thermal conduction and KHI. However, such an endeavor requires substantial computational resources. Nevertheless, we anticipate that the main results will remain consistent, grounded in considering the typical timescales of relevant mechanisms in our models, as detailed in Section 2.2.

\subsection{Caveats}

In this paper, we have implemented a simplified model to simulate the cold stream in a hot gaseous halo. Recognizing that several factors not included in our simulation may affect our results is necessary. First, our model does not account for the presence of central galaxies and their associated stars and interstellar medium, which will affect gas cooling in the central region. Moreover, feedback from central galaxies may have non-negligible effects on the fate of cold streams. For instance, \cite{2015MNRAS.448...59N} and \cite{2018MNRAS.478..255C} showed that feedback can strongly suppress the inflow rate of the gas at all redshifts, regardless of the temperature history of newly acquired gas. Second, halo gas is usually in a multi-phase state due to inflow, outflow, and thermal instability. Here, we have assumed that the halo gas is isothermal and in hydrostatic equilibrium. A more realistic model of the CGM, combined with improved resolution, would provide a more vivid picture of the evolution of cold stream (e.g.\citealt{2020ApJ...903...32F,2021MNRAS.507.2869S,2023ARA&A..61..131F}). Thirdly, throughout the simulation, the host dark matter halo and associated gravitational potential well are static. Considering the growth of dark matter mass as a function of time and merger, the strength of compressional heating will increase with time, potentially shortening the lifetime of cold streams. Furthermore, it is essential to recognize that our simulations exclude certain influential factors, such as magnetic fields and extrinsic turbulence, which could more or less influence the evolution of cold streams. The recent work by \cite{ledos2023stability} indicates that magnetic fields can potentially suppress the Kelvin-Helmholtz instability and facilitate cold streams to fuel galaxies. Therefore, for a more comprehensive investigation in the future, including magnetic fields in MHD simulations could provide valuable insights.

On the other hand, we assume a constant accretion rate of $3\, \rm{M_{\odot}}/yr$ for each cold stream onto halos with a mass of approximately $10^{12}\, \rm{M_{\odot}}$. This value is inferred from the results in cosmological simulations (e.g., \citealt{2008MNRAS.390.1326O, 2011MNRAS.414.2458V, 2022ApJ...924..132Z}). However, we did not incorporate the dependence of the cold accretion rate on the halo mass in this study. Some literature suggests a moderate increase in the cold accretion rate for halos with masses ranging from $10^{12}\, \rm{M_{\odot}}$ to $3\times 10^{12}\, \rm{M_{\odot}}$ (e.g., \citealt{2011MNRAS.414.2458V, 2022ApJ...924..132Z}). Cold stream with a higher accretion rate is expected to lift $M_{stream}$. However, the increase in cold accretion to more massive halos may be partly caused by more cold streams. A comprehensive study on the inflow rate of a single cold stream in halos with varying masses could offer a more accurate estimation of $M_{stream}$. Finally, it is noted that due to limitations in the version of Athena++ employed in this study, some modules conflict with the self-gravity of gas. The latest version of the Athena++ code has resolved this issue, allowing for the inclusion of the self-gravity of gas in future exploration. 

\section{Conclusions}

In this paper, we have performed a set of 3D hydrodynamic simulations to track the evolution of cold streams in hot gaseous halos supported by the gravitational potential of host dark matter halos with a mass between $10^{12}$ and $3.0 \times 10^{12}\, \rm{M_{\odot}}$ at $z=2$. The general evolution and the cooling and compressional heating rate of cold streams in the simulations have been probed. The accretion rate of cold and cool gas at various radii has been measured to estimate the survivability of cold streams quantitatively. We have explored the impact of dark matter halo mass, gas metallicity, simulation resolution, inhomogeneity, and thermal conduction on the fate of cold streams. Our findings are summarized as follows:

\begin{enumerate}
\item For a gas metallicity Z=0.1 $Z_{\odot}$, cold streams with mass injection rate $3 \,\rm{M_{\odot}}/yr$ per each can flow into the central region, i.e., $r<0.2\, r_{vir}$, of hot gaseous dark matter halo with mass $M_h = 10^{12} \,\rm{M_{\odot}}$ at $z=2$. In addition, cold streams induce runaway cooling in the central region. On the contrary, the cold streams could be substantially disrupted by the heating due to compression at a radius around 0.2  $r_{vir}$ in a more massive halo with $M_h = 1.5$ and $3.0 \times 10^{12} \,\rm{M_{\odot}}$. 

\item The rivalry between cooling and compressional heating is the decisive factor governing the fate of the cold stream in the hot gaseous halo, which is similar to the case of spherical accretion. In the case of a $M_{h}=1\times 10^{12}\, \rm{M_{\odot}}$ halo with a gas metallicity Z=0.1 $ Z_{\odot}$, the cooling/compression ratio of a mixed gas of the cold stream and the hot halo can reach 1000, which means a pronounced dominance of cooling. As a result, the cold stream can be sustained to the central region of the halo. However, for a $M_{h}=3\times 10^{12} \rm{M_{\odot}}$ halo, the cooling rate of mixed gas either marginally balances with or falls below the compressional heating rate. Consequently, the cold accretion via stream into the central region of the halo is suppressed.

\item The transition halo mass, $M_{stream}$, which determines whether cold streams can persist or not, increases with increasing gas metallicity. Higher metallicity increases the cooling rate of gas, thereby enhancing the likelihood of cold stream survival. For Z=0.001 $Z_{\odot}$, the transition mass is in the range of 1.0 and $1.5\times 10^{12}\,\rm{M_{\odot}}$. If the metallicity increases to Z=0.1 $Z_{\odot}$, the transition mass increases to approximately $1.5\times\,\rm{10^{12}M_{\odot}}$. These values are generally consistent with previous theoretical studies of spherical accretion and recent observations (e.g., \citealt{2022ApJ...926L..21D}). 

\item In halos less massive than the transition mass, $M_{stream}$, cold streams can efficiently deliver cold and cool gas to the central region through two mechanisms. The first is the gas within the cold streams that initially entered from outside the halo. The second mechanism arises from the cooling of hot gas within the central region when it mixes with the injected cold gas. In some extreme cases, with the runaway cooling, the latter can dominate over the former. For instance, in the simulation with $M_{h}=1\times 10^{12}\, \rm{M_{\odot}}$, the mass growth of cold and cool gas within $r=0.2r_{vir}$ can be much higher than the inflow rate at the sphere of $r=0.2r_{vir}$ during some stages. This suggests that the cold and cool gas within $r=0.2r_{\text{vir}}$ is mainly sourced from the gas already within that same radius rather than from gas external to this region. Nevertheless, the second mechanism warrants further validation through a more realistic model of the central region of the halo, particularly including a central galaxy and its feedback.
\end{enumerate}

In summary, our investigation provides insights into the mechanisms and processes that disrupt cold streams in intermediate-mass halos.

\section*{Acknowledgements}

We thank the anonymous referee for her/his very useful comments and suggestions that improved the manuscript. This work is supported by the National Science Foundation of China (Nos. 11733010, 12173102, 11833005, 11890692), 111 project No. B20019, and Shanghai Natural Science Foundation, grant No. 19ZR1466800. We acknowledge the science research grants from the China Manned Space Project with Nos. CMS-CSST-2021-A02. Simulations carried out in this work were completed on the HPC facility of the School of Physics and Astronomy, Sun Yat-Sen
University.

\section*{Data Availability}

The data underlying this article will be shared with the corresponding author under reasonable request.



\bibliographystyle{mnras}
\bibliography{example} 



\appendix
\section{Impact of inhomogeneity in cold streams }

\begin{figure*}
    \centering
    \includegraphics[trim ={1.5cm 2.0cm 2.cm 0cm},clip,width=2.0\columnwidth]{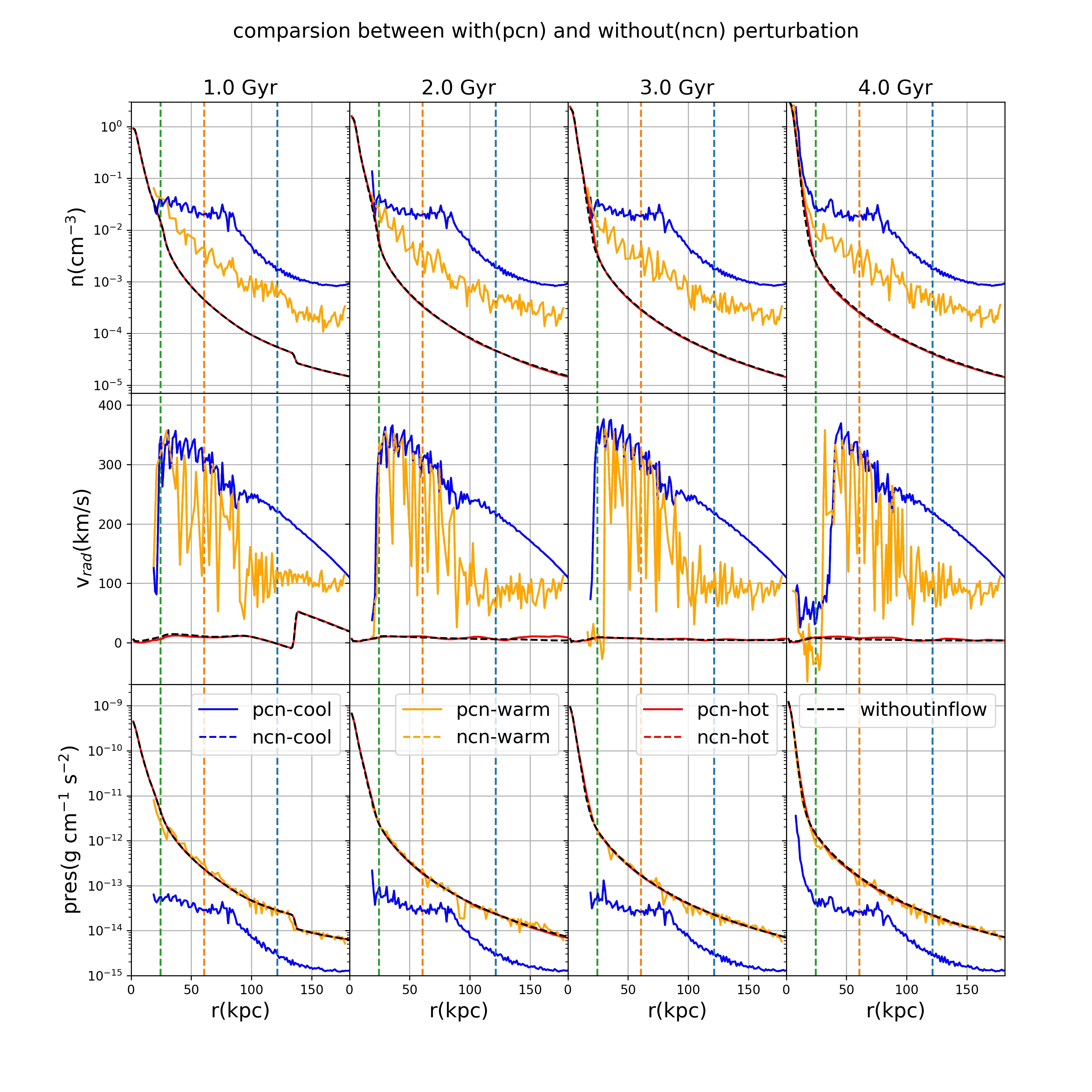}
    \caption{The same as Figure \ref{fig:m10ma10-Curve}, but for the simulation 1e12-256-pcn and 1e12-256-ncn, in which $M_h = 1 \times 10^{12}\,\rm{M_{\odot}}$. The solid line is overlaid with the dotted line.}
    \label{fig:perturbation-compare}
\end{figure*}

Previous studies have demonstrated that the IGM exhibits inhomogeneities. The simulations in previous sections assume the density in the cold stream is non-uniform due to the inhomogeneous IGM. We performed one simulation without density perturbation in the injected cold gas to investigate the impact of non-uniform density distributions in the cold streams on their evolution. The physical properties presented in Figure \ref{fig:perturbation-compare} offer a comparison between two simulations with/without inhomogeneities in cold streams for a halo mass $10^{12}\,\rm{M_{\odot}}$ and a gas metallicity Z=0.001 $Z_{\odot}$. Solid lines and dotted lines are overlaid with the same color. Inhomogeneity does not affect the final result of the simulation. However, cold streams with non-uniform density distribution should be more realistic.

\section{Comparison with a high-resolution run}
\label{sec:appd-resolution}

\begin{figure*}
    \centering
    \includegraphics[trim ={1.5cm 2.0cm 2.cm 0cm},clip,width=2.0\columnwidth]{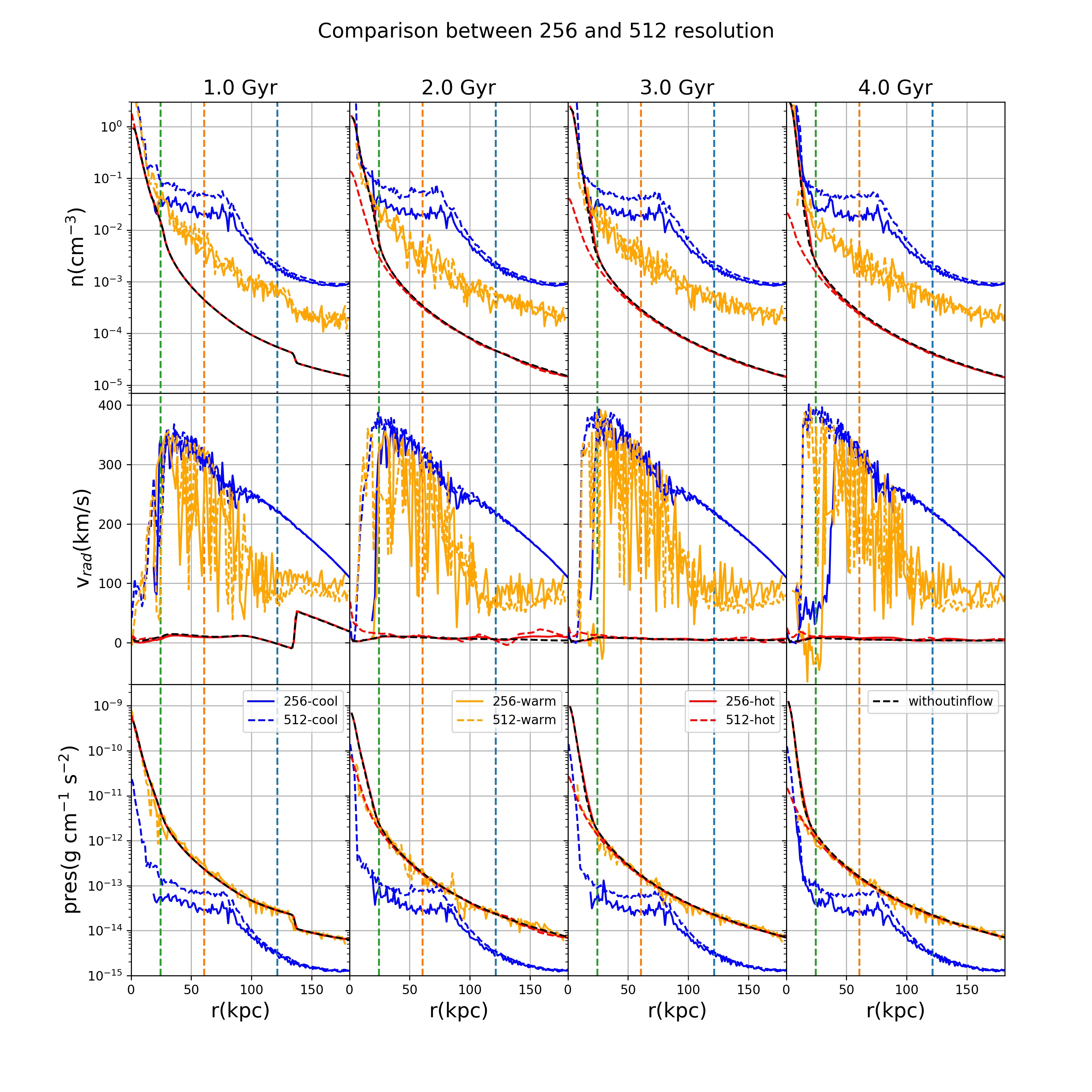}
    \caption{The same as Figure \ref{fig:m10ma10-Curve}, but for the simulation 1e12-256-pcn and 1e12-512-pcn, in which $M_h = 1 \times 10^{12}\, \rm{M_{\odot}}$. The solid line is the result of 256 resolution, the colored dashed line is the result of 512 resolution, and the black dashed line is the control result of no cold inflow.}
    \label{fig:res-compare}
\end{figure*}

To save computing resources, we have adopted a default grid number of $256^3$. To gauge the impact of resolution, we have conducted a pair of simulations, i.e., `1e-256-pcn' and `1e-512-pcn', with a grid size of $256^3$ and $512^3$, respectively, for a halo mass of $10^{12}\,\rm{M_{\odot}}$ and a metallicity of $z=0.001 Z_{\odot}$. To provide a quantitative comparison, Fig. \ref{fig:res-compare} shows the physical properties of these two simulations that only differ in resolution. The solid lines represent the simulation with $256^3$ grids, while the dotted lines represent the simulation with $512^3$ grids. The profiles at $r>0.2\, r_{vir}$ are similar between the two simulations over time. In both simulations, the cold stream reaches $r=0.2\,r_{vir}$ at $t\sim 1.0$ Gyr, and there is an accumulation of cold gas in the central region at $t\sim 4.0$ Gyr. Yet, the accumulation of cold gas in the higher resolution run starts earlier. Meanwhile, cold gas can flow into the central region of the halo with a larger maximum velocity in the higher resolution run after $t\sim 1.0$ Gyr. In short, the overall evolution of the cold stream at $r \gtrsim 0.2 \, r_{vir}$ converges at a resolution of $256^3$. Meanwhile, a resolution higher than $256^3$ is urged when considering the state of gas in the very central region of the halo (r<10 kpc). It is worth noting that our current model excludes the central galaxy. Accordingly, in addition to higher resolution, a more realistic model of the central galaxy is required to obtain more reliable results on the state of gas in the central region. 

\section{Results of the simulation run 1e12-256-pcn}
\label{sec:appdenix-pcn}

\begin{figure*}
    \centering  \includegraphics[trim ={1.5cm 2.0cm 2.cm 0cm},clip,width=2.0\columnwidth]{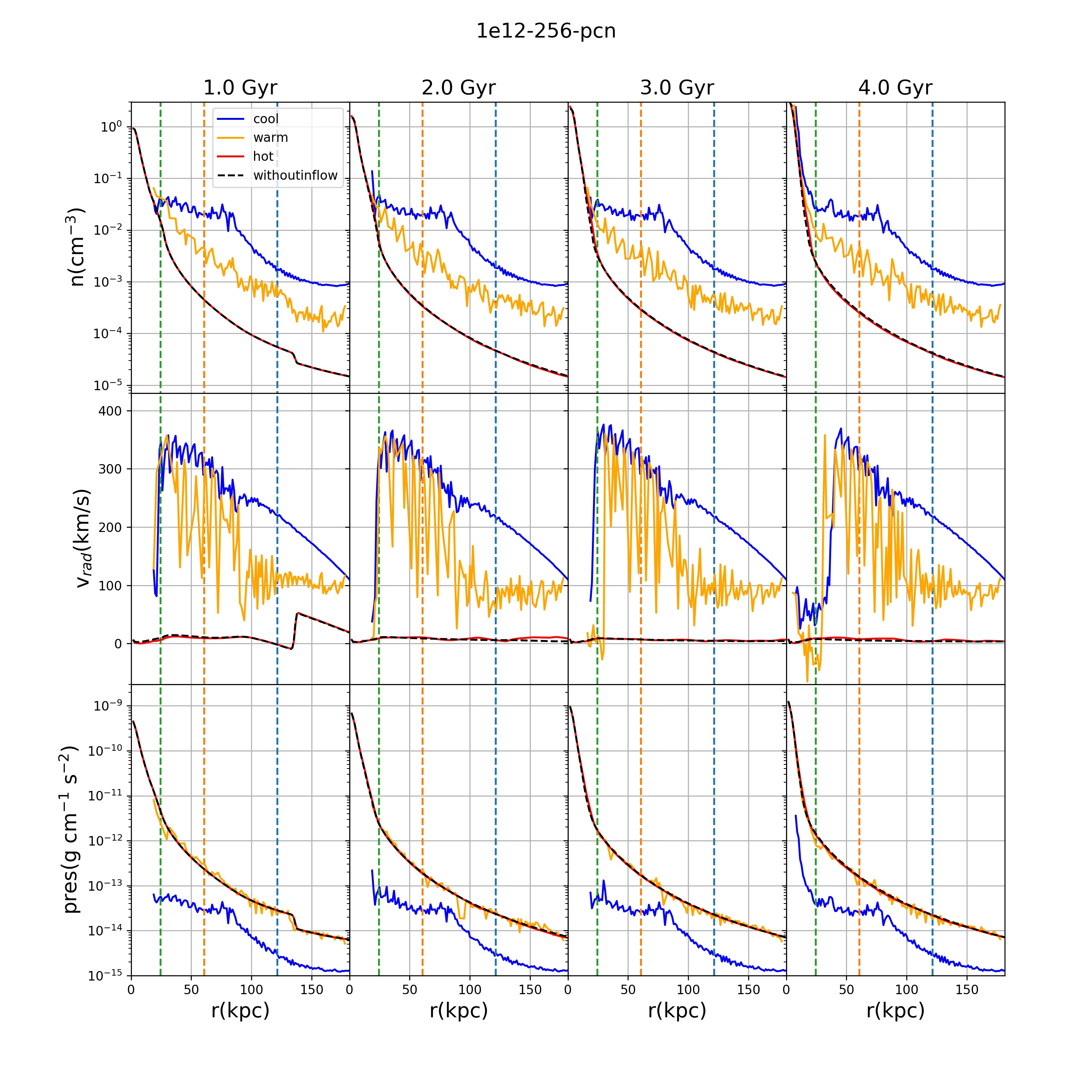}
    \caption{Rows from the top to bottom show the averaged number density, radial velocity, and pressure, as functions of distance from the halo center respectively for the simulation 1e12-256-pcn, in which $M_h=10^{12} \,\rm{M_{\odot}}$ and $Z=0.001Z_{\odot}$. Results at times 1.0, 2.0, 3.0, and 4.0 are presented from the left to the right columns, respectively. The red, orange, and blue lines indicate hot (T$\geq 10^{6}$ K), warm ($10^{5}  \leq T < 10^{6}$ K), and cold and cool gas ($ T < 10^{5}$ K). The black dashed lines are the simulation results with the same conditions but without a cold stream.} Vertical dotted lines indicate r= 0.2 (green), 0.5 (orange), 1.0 (blue) $r_{vir}$.
    \label{fig:m30ma10-Curve}
\end{figure*}

\begin{figure*}
    \centering
    \includegraphics[trim = {3.2cm 2cm 4cm 0cm},clip,width=2.0\columnwidth]{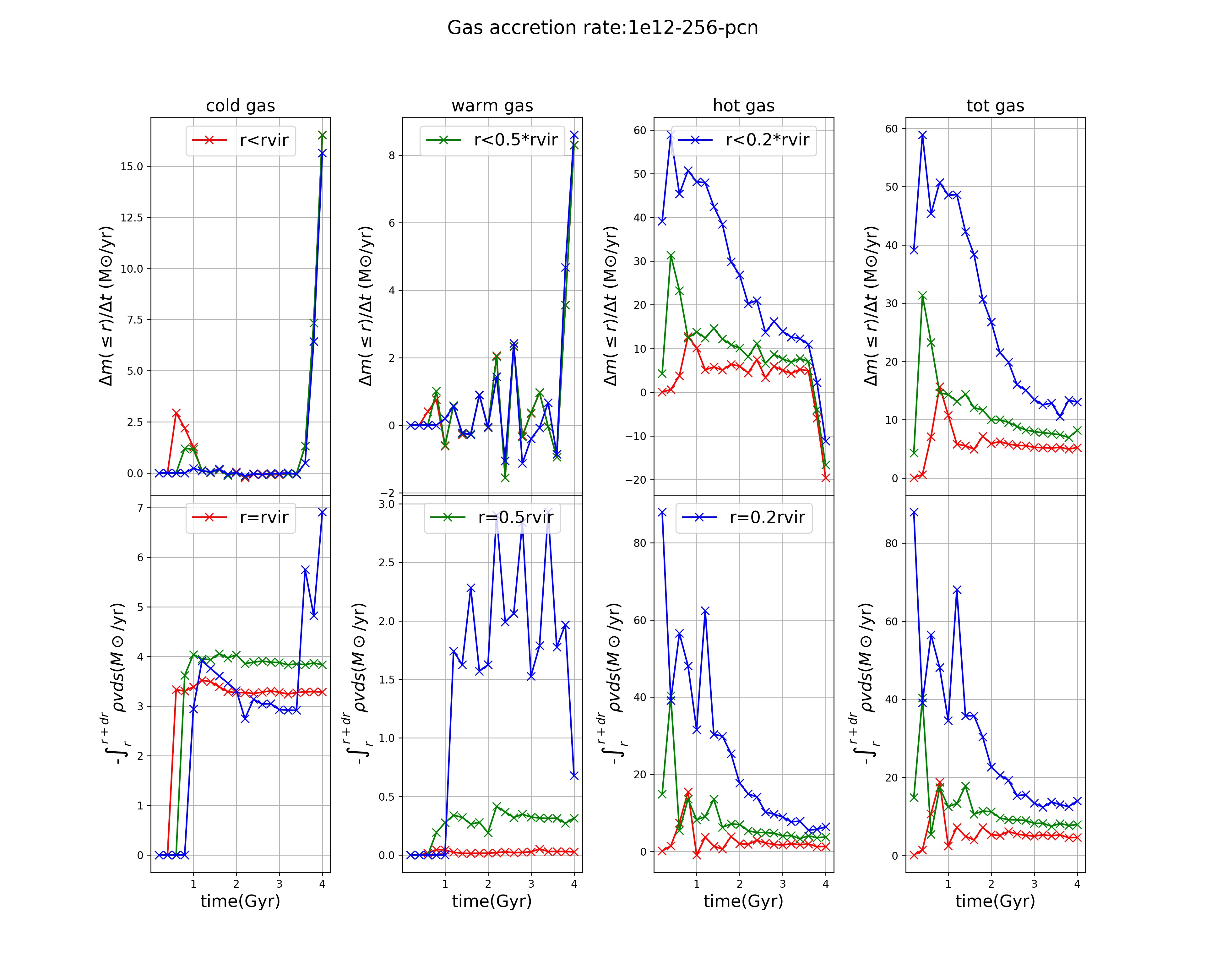}
    \caption{Gas accretion rate for the simulation 1e12-256-pcn, in which $M_h=10^{12}\, \rm{M_{\odot}}$ and $Z=0.001Z_{\odot}$. The upper row indicates $\bar{\dot{M}}(r,t)$ within r=1.0$r_{vir}$(red), r=0.5$r_{vir}$(green) and r=0.2$r_{vir}$(blue). The bottom row presents the instant accretion rate $\dot{M}(r,t)$ at r=1.0$r_{vir}$, r=0.5$r_{vir}$ and r=0.2$r_{vir}$. See Equations \ref{eqn:ave_accretion} and \ref{eqn:ins_accretion} for the definition of accretion rate. From the left to the right column, the plots show the cold and cool, warm, hot, and all the gas, respectively.}
    \label{fig:m30ma10-accre}
\end{figure*}

\begin{figure*} 
    \centering
    \includegraphics[trim ={2cm 4cm 2cm 0cm},clip, width=2.0\columnwidth]{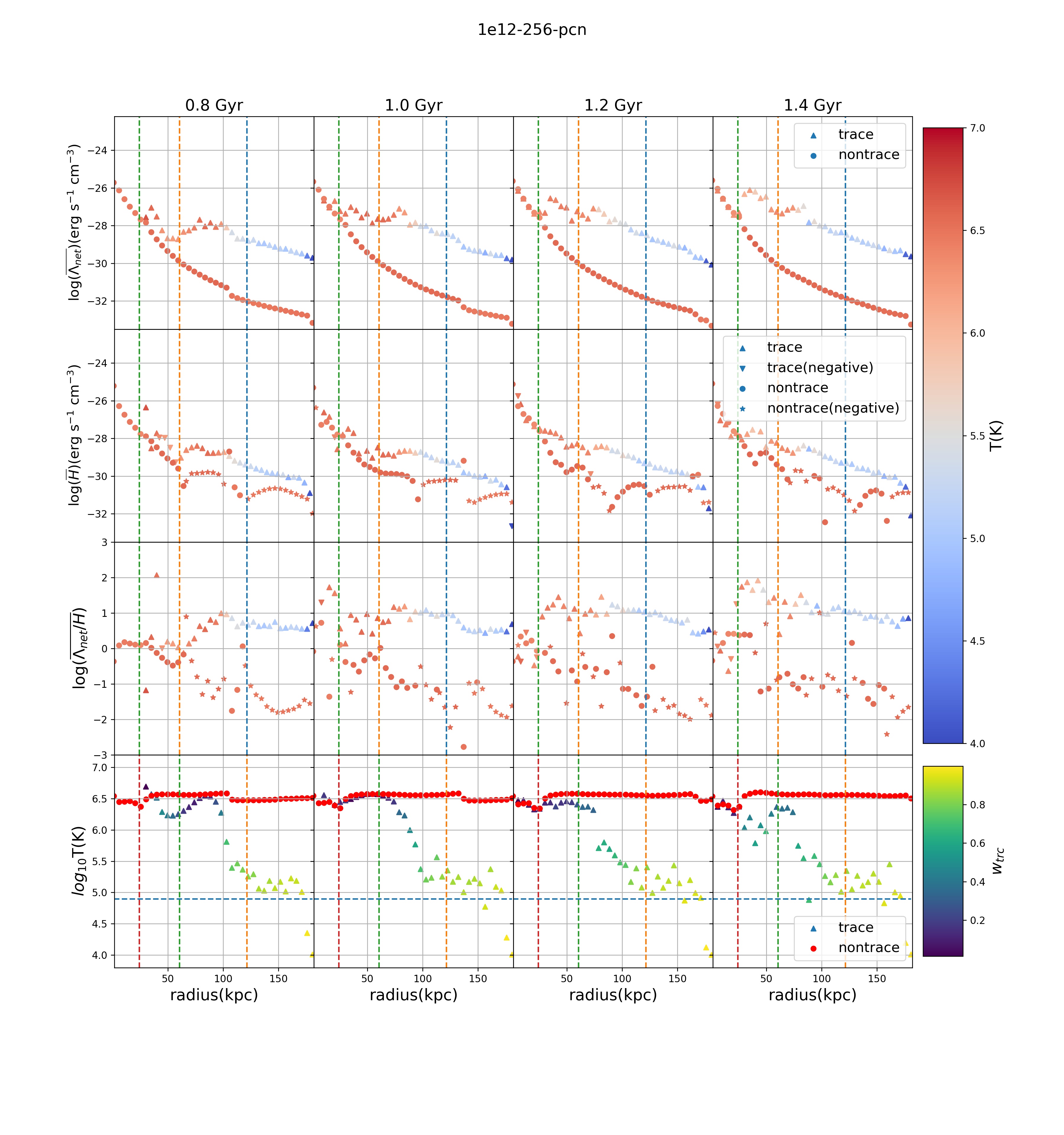}
    \caption{The first, second, third, and fourth column show the mean cooling, compression rate, ratio of cooling/compression, and the average temperature as functions of the radius from halo center, respectively, at time 0.8, 1.0, 1.2, 1.4 Gyr (from top to bottom) for the simulation 1e12-256-pcn, in which $M_h = 10^{12}\, \rm{M_{\odot}}$ and $Z=0.001Z_{\odot}$. The data points in the first three columns are color-coded by the average temperature. The triangles (circles) represent the average values of gas cells (do not) containing tracers of injected cold streams. In the second column, the downward triangles (stars) indicate that the gas cells (do not) contain a tracer of injected cold streams and have a negative compression rate. The vertical dotted lines indicate r= 0.2 (green), 0.5 (orange), 1.0 (blue) $r_{vir}$.}  
    \label{fig:m30ma10-lamb}
\end{figure*}

In the main text, we have shown the results of simulations with a higher metallicity of $Z=0.1\,Z_{\odot}$, which could provide a sharper view of the transition of the final fate of the cold stream when the halo mass increases. For simulations with a lower metallicity $Z=0.001\, Z_{\odot}$, the transition is less evident in visual impression.  As a reference, we would like to briefly show the results of the simulation run `1e12-256-pcn', which has a halo mass of 1.0 $\times 10^{12}\,\rm{M_{\odot}}$ and a gas metallicity of $Z=0.001 Z_{\odot}$ below. Figure \ref{fig:m30ma10-Curve}, \ref{fig:m30ma10-accre}, and \ref{fig:m30ma10-lamb} present the radial profiles, accretion rates, and cooling and compression for the halo mass 1.0 $\times 10^{12}\,\rm{M_{\odot}}$ with metallicity $Z=0.001 Z_{\odot}$. Compared to the simulation run `1e12-256-pcn-Hz' (Figure \ref{fig:m10ma10-Curve}, \ref{fig:m10ma10-lamb}, \ref{fig:m10ma30-accre}), the accumulation of cold gas in the central region is delayed and suppressed significantly in `1e12-256-pcn'. 

\bsp	
\label{lastpage}
\end{document}